\documentclass[conference]{hpca23-templete/IEEEtran}
\usepackage{cite}
\usepackage{amsmath,amssymb,amsfonts}
\usepackage{algorithmic}
\usepackage{graphicx}
\usepackage{textcomp}
\usepackage{xcolor}
\usepackage[hyphens]{url}
\usepackage{tablefootnote}
\usepackage{pifont}

\usepackage{comment}
\usepackage{algorithm2e}
\usepackage{fancyhdr}
\usepackage[bookmarks=true,breaklinks=true,colorlinks,linkcolor=blue,citecolor=blue,urlcolor=black]{hyperref}
\usepackage{listings}
\usepackage[normalem]{ulem}

\def\BibTeX{{\rm B\kern-.05em{\sc i\kern-.025em b}\kern-.08em
    T\kern-.1667em\lower.7ex\hbox{E}\kern-.125emX}}
\newcommand{\tileload}{\textsc{tile\_load}\xspace}
\newcommand{\tileloadt}{\textsc{tile\_load\_t}\xspace}
\newcommand{\tileloadu}{\textsc{tile\_load\_u}\xspace}
\newcommand{\tileloadv}{\textsc{tile\_load\_v}\xspace}

\newcommand{\tilestore}{\textsc{tile\_store}\xspace}
\newcommand{\tilestoret}{\textsc{tile\_store\_t}\xspace}

\newcommand{\tilemm}{\textsc{tile\_gemm}\xspace}

\newcommand{\matmul}{\textsc{tile\_gemm/tile\_spmm}\xspace}

\newcommand{\metaload}{\textsc{tile\_load\_m}\xspace}
\newcommand{\tilespmmu}{\textsc{tile\_spmm\_u}\xspace}
\newcommand{\tilespmmv}{\textsc{tile\_spmm\_v}\xspace}

\newcommand{\tilespmmr}{\textsc{tile\_spmm\_r}\xspace}

\newcommand{\GJ}[1]{{\color{brown}\bfseries [Geonhwa:: #1]}}

\newcommand{\HK}[1]{{\color{violet}\bfseries [Hyesoon:: #1]}}

\newcommand{\TODO}[1]{\textcolor{red}{TODO:: #1}}
\newcommand{\rev}[1]{\textcolor{black}{#1}}

\def\Snospace~{\S{}}

\title{VEGETA: \underline{V}ertically-Integrated \underline{E}xtensions for Sparse/Dense \underline{GE}MM \underline{T}ile \underline{A}cceleration on CPUs}

\author{
    \IEEEauthorblockN{Geonhwa Jeong\IEEEauthorrefmark{1}, Sana Damani\IEEEauthorrefmark{1}\textsuperscript{\textsection}, Abhimanyu Rajeshkumar Bambhaniya\IEEEauthorrefmark{1}, Eric Qin\IEEEauthorrefmark{1}\textsuperscript{\textsection},  \\
    Christopher J. Hughes\IEEEauthorrefmark{2}, Sreenivas Subramoney\IEEEauthorrefmark{2}, Hyesoon Kim\IEEEauthorrefmark{1}, and Tushar Krishna\IEEEauthorrefmark{1} 
    \IEEEauthorblockA{
        \IEEEauthorrefmark{1}\textit{Georgia Institute of Technology}, 
        \IEEEauthorrefmark{2}\textit{Intel Labs} \IEEEauthorblockA{\\
        \IEEEauthorrefmark{1}\{geonhwa.jeong, sdamani, abambhaniya3, ecqin\}@gatech.edu,
        hyesoon@cc.gatech.edu, tushar@ece.gatech.edu}
        \IEEEauthorrefmark{2}\{christopher.j.hughes, sreenivas.subramoney\}@intel.com
        }
    }
}

\begin{document}
\maketitle
\begingroup\renewcommand\thefootnote{\textsection}
\footnotetext{Sana Damani and Eric Qin are now at NVIDIA and Meta, respectively.}
\endgroup

\thispagestyle{plain}
\pagestyle{plain}


\begin{abstract}
Deep Learning (DL) acceleration support in CPUs has recently gained a lot of traction, with several companies (Arm, Intel, IBM) announcing products with specialized matrix engines accessible via GEMM instructions.
CPUs are pervasive and need to handle diverse requirements across DL workloads running in edge/HPC/cloud platforms. 
Therefore, as DL workloads embrace sparsity to reduce the computations and memory size of models,
it is also imperative for CPUs to add support for sparsity to avoid under-utilization of the dense matrix engine and inefficient usage of the caches and registers. 
This work presents VEGETA, a set of ISA and microarchitecture extensions over dense matrix engines to support flexible structured sparsity for CPUs, enabling programmable support for diverse DL models with varying degrees of sparsity. 
Compared to the state-of-the-art (SOTA) dense matrix engine in CPUs, a VEGETA engine provides 1.09$\times$, 2.20$\times$, 3.74$\times$, and 3.28$\times$ speed-ups when running 4:4 (dense), 2:4, 1:4, and unstructured (95\%) sparse DNN layers.


\end{abstract}
\section{Introduction}
Deep learning (DL) is used in various domains including computer vision, recommendation systems, and natural language processing~\cite{mlfb_hpca19, dlrm19, tpu-isca}. 
However, the training and inference for those DL models require a large number of \textbf{computations} and huge \textbf{memory}. 
To accelerate those, high-end GPUs (such as ones with NVIDIA Tensor cores~\cite{nvidia_volta}), domain-specific accelerators (such as Google's TPUs~\cite{tpu-isca}), and high-performance CPUs~\cite{intel2018vnni, intel2020isa} along with optimized libraries and frameworks~\cite{onednn, tvm, cudnn, tensorflow2015-whitepaper} have been introduced.

Although CPUs have received relatively less attention in the DL era compared with GPUs and domain-specific accelerators, CPUs are widely used at datacenters thanks to their flexibility~\cite{mlfb_hpca19}. 
In fact, there are cases where CPUs are more suitable than GPUs/accelerators as the primary processor for processing deep neural networks (DNNs). 
For example, edge devices with tight area/power budget~\cite{sparce_taco19} prefer CPUs since GPUs/accelerators cannot be easily deployed in addition to CPUs. 
Moreover, a large number of server-class CPUs are (and will be) used in datacenters to process High Performance Computing (HPC) workloads, and adding/maintaining additional accelerators just for DL workloads would increase complexity and cost~\cite{save_micro20}. 
Furthermore, offloading a modest-size task to an accelerator or a GPU might not be the best option if the offloading overhead 
is relatively substantial~\cite{accelerometer}.
Finally, large DNNs are hard to be used with GPUs or accelerators due to their smaller size of memory compared with that of CPUs~\cite{intel-mlperf}. 
Considering these reasons, it is not surprising that many CPU vendors including Arm~\cite{arm20ethos}, Intel~\cite{intel2020isa}, and IBM~\cite{ibm21micro} have started deploying \textit{dense} matrix engines along with conventional scalar and vector engines to accelerate GEMM (i.e., general matrix multiplication) which is at the core of DL models~\cite{gemm_petewarden}. 
Since they only focus on dense matrix computations, there is a challenge as sparsity becomes pervasive in DL. Thus, it is time to ponder how to utilize these engines efficiently for sparse DNNs on CPUs, similar to how we have evolved scalar/vector engines over many years. 



Recent research has focused on adding hardware support in DNN accelerators to handle sparsity as it can be used to improve power and performance by skipping or gating ineffectual computations (since anything multiplied by zero is zero) and to reduce 
memory capacity and bandwidth requirements by saving only non-zero values and their indices in a compressed manner~\cite{extensor, sigma, scnn, eyeriss_v2}.
CPUs, however, bring in the following additional challenges.
(i) Being general-purpose, CPUs are particularly sensitive to the amount of hardware devoted to improving a subset of workloads.
(ii) At the same time, new functional units in CPUs need to be able to support a wide variety of use cases.
(iii) CPUs need programmability support to handle sparsity unlike offload accelerators with custom DMA engines for sparse accesses.


In this paper, we introduce VEGETA, \underline{V}ertically-Integrated \underline{E}xtensions for Sparse/Dense \underline{GE}MM \underline{T}ile \underline{A}cceleration on CPUs. 
To expose VEGETA to programmers, we extend the ISA with new instructions and registers, which enables accelerating sparse DNNs by removing redundant computations and reducing memory traffic. 
We also introduce a new light-weight matrix engine that enhances a systolic array and register file to support flexible $N$:$M$ structured sparsity.
Further, we demonstrate how VEGETA can accelerate \textit{both} structured and unstructured sparse-dense GEMMs (SPMMs) via software transformations of the sparse matrix.
This is the first work, to the best of our knowledge, to demonstrate sparsity support in a CPU matrix engine.
We believe this work is extremely timely given growing interest in industry products for sparsity, such as Sparse Tensor Core (STC) in NVIDIA's Ampere GPU~\cite{mishra2021accelerating} and Sparsity-aware NPU in a Samsung Mobile AP~\cite{samsung}.

\textbf{Summary of Contributions:}

\begin{itemize}
\item We introduce VEGETA ISA extensions that include new instructions and registers for supporting sparse tile operations along with software optimizations.
\item We propose new architectural extensions to a systolic-array-based matrix engine to support flexible $N$:$M$ structured sparsity in terms of sparsity pattern and granularity.
\item We explore different VEGETA engine design choices to understand the performance and area trade-offs.
\item Across a suite of GEMM/SPMM kernels, we observe 1.09$\times$, 2.20$\times$, 3.74$\times$, and 3.28$\times$ speed-ups compared to a SOTA dense array when running 4:4 (dense), 2:4, 1:4, and unstructured (95\%) sparse DNN layers, respectively. 
\end{itemize}
\section{Background}
\subsection{GEMM and DNN Layers}
A DNN comprises multiple layers of potentially different types. 
Two of the most time-consuming layer types are fully connected layers, which are widely used in Multi-Layer Perceptrons/Transformers, and convolutional layers~\cite{eyeriss}.
The underlying building block of these popular and compute-intensive DNN layers is GEMM.
Fully connected layers are composed of GEMMs naturally, while modern high-performance convolutional layers are implemented with GEMMs as their innermost operation~\cite{libxsmm_ipdps20}.

\subsection{Systolic Array (SA)}
In the last few years, systolic arrays~\cite{systolic} have become a prime design choice for accelerating dense GEMM, especially in the context of DNNs, due to its simple and efficient architecture.
The most well-known commercial systolic array architecture is in Google's TPU~\cite{tpu-isca}.
A systolic array is a two-dimensional array composed of highly scalable, pipelined, and homogeneous processing elements (PEs). 
Each PE is responsible for a Multiply-and-Accumulate (MAC) operation and forwarding operands to the neighboring PEs. 
An element of one matrix is pre-loaded into the PE (often called the stationary element~\cite{eyeriss}).  
The elements of the remaining matrices are streamed to the left and top edge of the systolic array. Then, they are propagated to the PEs using the horizontal and vertical links between neighboring PEs.  
This maximizes data reuse between PEs, thus minimizing the number of reads and writes between the array and external buffers.


\subsection{Sparsity in DNNs}

\begin{figure}[!t]
    \centering
    \includegraphics[width=0.5\textwidth]{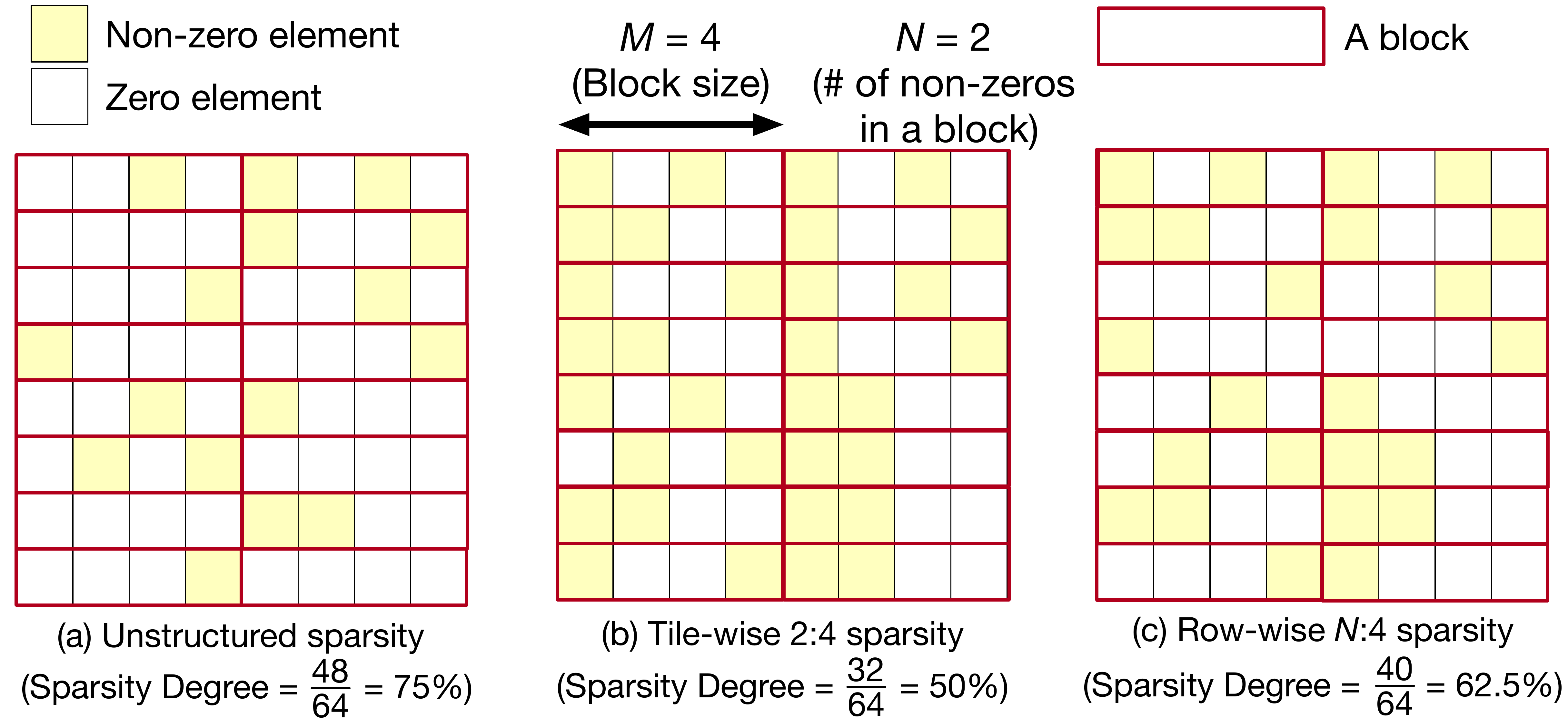}
    \vspace{-1em}
    \caption{
    Comparison with unstructured sparsity, tile-wise 2:4 sparsity and row-wise $N$:4 sparsity (different rows could have different $N$ values between 0 and 4). The size of tile is 8$\times$8 in this example and the size of a block is 4.
    }
    \label{fig:sparsity}
\end{figure}

\begin{figure}[!t]
    \centering
    \includegraphics[width=0.45\textwidth]{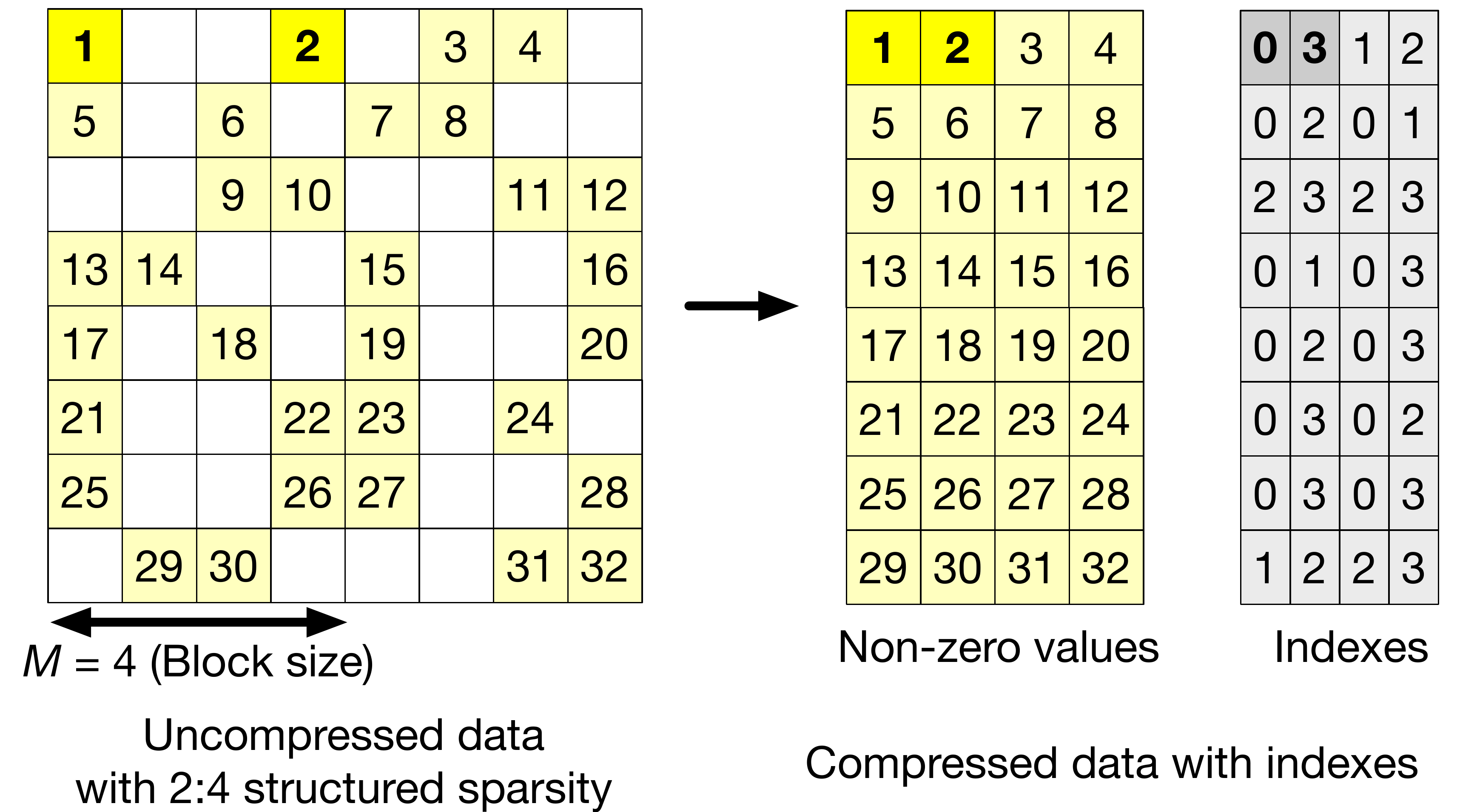}
    \caption{Compression of a matrix with $N$:$M$ structured sparsity. The $M$ determines the bit widths of indexes. 
    }
    \label{fig:compressed-matrix}
    \vspace{-1em}
\end{figure}
Even though the number of computations required for executing a DNN is huge, there is an opportunity to reduce that significantly by leveraging sparsity in DNNs. We describe different dimensions of sparsity in DNN inference. 

\textbf{Sparsity source.} Weights and input activations are the main sources of sparsity in DNNs.
Weight (static) sparsity is derived by pruning some edges in a DNN with a small sacrifice in accuracy~\cite{pruning_nips15, deepcompression_iclr16}.  
This leads to zeros in the weight matrix.
Input (dynamic) sparsity is usually induced by a popular activation function, Rectified Linear Unit (ReLU), which clips negative inputs to zero, leading to a significant fraction of zero-valued input activations for the next layer.

\textbf{Sparsity degree.} Sparsity degree is the fraction of zeros in a given matrix. 
The degree of weight sparsity can often be tuned during the pruning process. However, the degree of input sparsity is usually non-deterministic and unpredictable since the actual values of input activations are determined at runtime.

\textbf{Sparsity pattern.}
The non-zeros for a DNN may have a pattern, such as certain channels of weights being all zero (through channel pruning) or each block with M elements having at most N non-zeros (often called $N$:$M$ sparsity~\cite{domino2021nips}).
``Unstructured sparsity'' indicates the lack of a pattern. 

\textbf{Sparsity granularity.}
Sparsity patterns may exist at different granularities. 
For example, ``network-wise $N$:$M$ sparsity'' indicates all layers in a network have the same $N$:$M$ ratio, while ``layer-wise $N$:$M$ sparsity'' means different layers may have different $N$:$M$ ratios.
During execution, usually a layer is decomposed into 2D tiles; each tile is composed of rows of vectors. 
Similar to the network granularity as explained before, sparsity patterns could exist at the tile or row granularity.
In \autoref{fig:sparsity}, we compare matrices with different sparsity patterns/granularities, with a tile size of 8$\times$8.
Previous HW support focused on either network-wise~\cite{mishra2021accelerating, sta_cal20, maohua_micro19} or layer-wise $N$:$M$ sparsity~\cite{s2ta_hpca22}, while we target to support row-wise $N$:$M$ sparsity to cover broader flexibility as well as some unstructured sparsity as shown in \autoref{table:nm-hw}.

\section{Motivation and VEGETA Overview}


\subsection{Vector vs. Matrix Engine}
\label{sec:vector_matrix}
\begin{figure}[!t]
    \centering
    \includegraphics[width=0.48\textwidth]{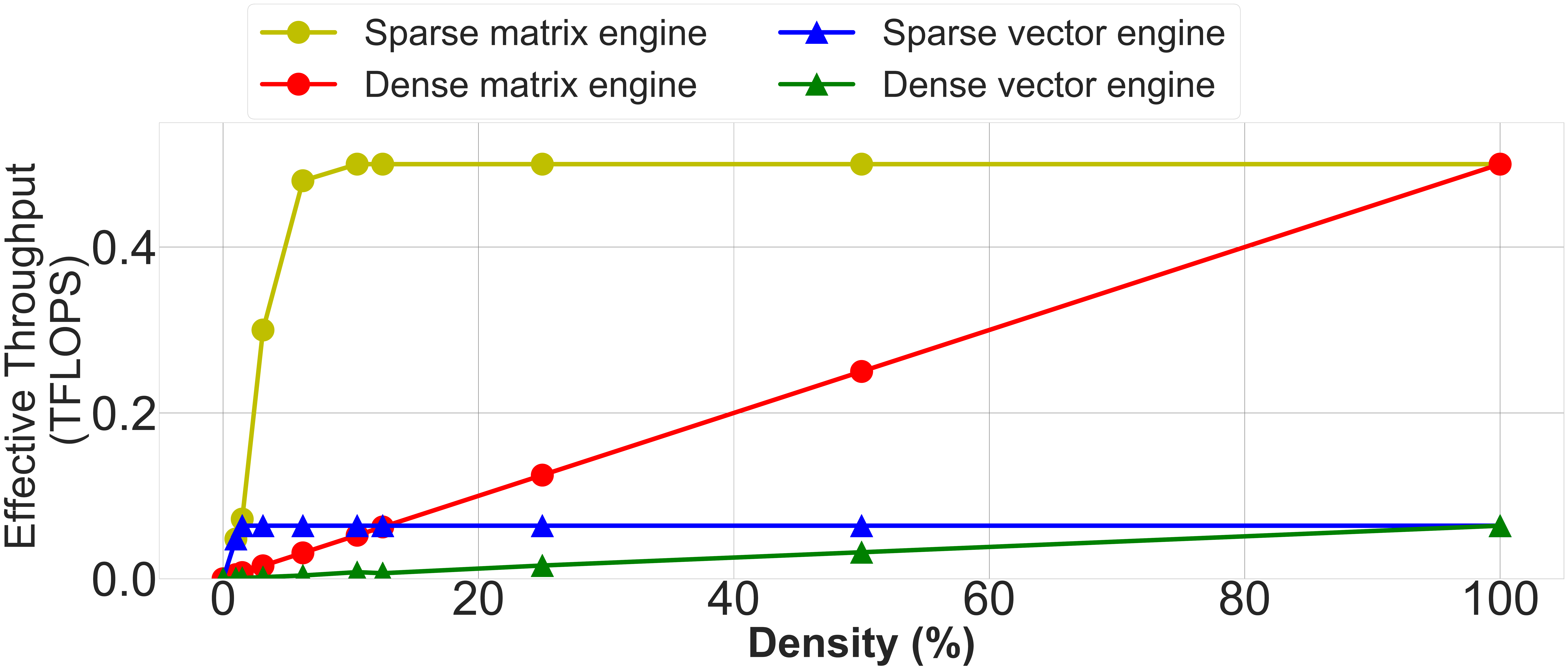}
    \vspace{-1em}
    \caption{Effective compute throughput for dense/sparse vector/matrix engines using a roofline model.}
    \label{fig:vector-matrix-comparison}
\end{figure}

\begin{figure}[!t]
    \centering
    \includegraphics[width=0.48\textwidth]{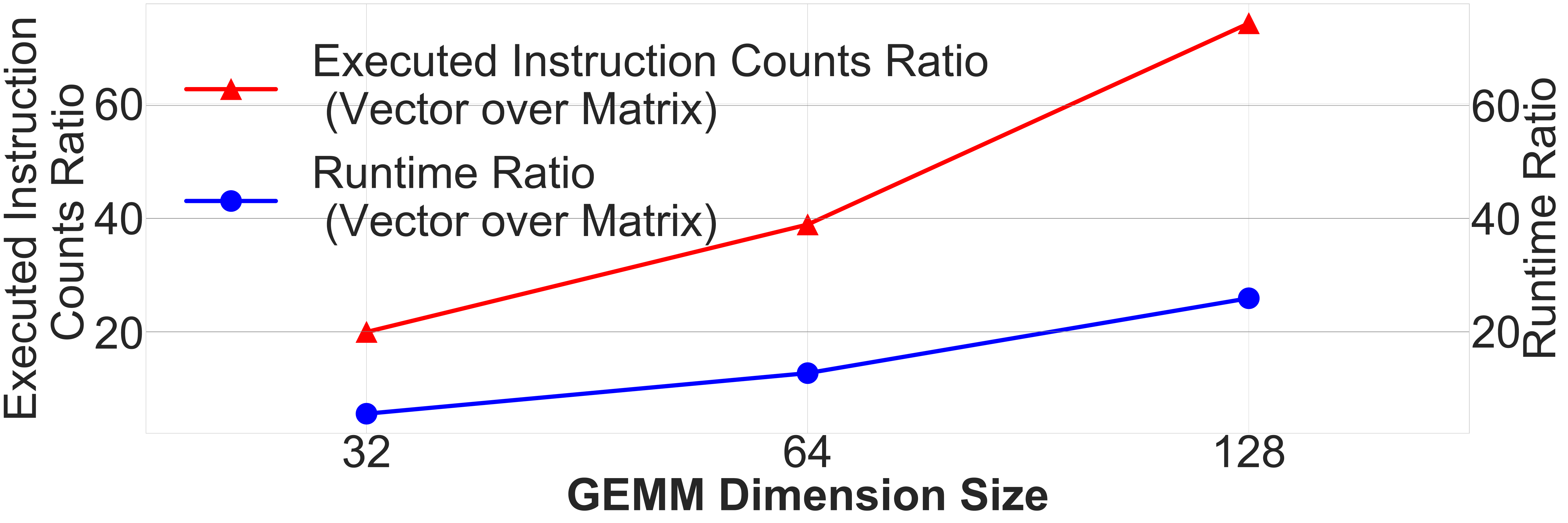}
    \vspace{-1em}
    \caption{Executed instruction counts and runtime ratio comparison on a CPU with matrix engines on GEMM workloads with equal-sized dimensions.
    }
    \label{fig:skx-spr-comparison}
\end{figure}

Single Instruction Multiple Data (SIMD) and/or short vector execution have been supported in mainstream CPUs for several generations. 
Due to the smaller granularity of vectors compared to matrices, the industry has started integrating matrix engines inside CPUs~\cite{ intel2020isa, arm20ethos, ibm20isa}. 
A matrix engine can provide power-efficient high compute throughput via simple control logic and significant internal data reuse. 
For example, in Intel's upcoming Sapphire Rapids processors, the peak matrix compute capability per cycle of each core is 8$\times$ the vector capability~\cite{intel2021}. 
In \autoref{fig:vector-matrix-comparison}, we show effective compute throughputs of sparse/dense matrix/vector engines on a convolutional layer with different densities derived from a roofline model.
We assume 64 and 512 GFLOPS for the vector and matrix engines, respectively, with a memory bandwidth of 94 GB/s~\cite{intelBWcal_orig}.
For the 100\% dense case, the dense matrix (vector) and sparse matrix (vector) engines achieve the same compute throughput since no computation can be skipped.
We observe that sparse engines outperform dense engines significantly by skipping non-effectual computations, especially when density is low.
Also, there is a significant gap in compute throughput between vector and matrix engines.
Moreover, due to the smaller granularity of vector instructions, the same GEMM kernel requires many more instructions to be executed when using vector engines, contributing to the runtime gap as shown in \autoref{fig:skx-spr-comparison} (we estimated them using a cycle accurate simulator, MacSim~\cite{hyesoon12macsim}).
When memory bound, i.e., at extremely low density and thus arithmetic intensity, vector compute throughput is sufficient, so a sparse vector engine performs similar to a sparse matrix engine.


\subsection{Structured vs. Unstructured Sparsity} 

CPUs are general-purpose processors. Ideally, a CPU should be able to support {\it any} sparsity pattern, including unstructured sparsity. However, this introduces two practical challenges.

\textbf{Challenge 1: Programmability.} 
GEMM implementations typically partition the matrices into tiles and iterate over the tiles to optimize data reuse in caches, facilitate parallelization, etc. Innermost GEMM kernels are often optimized for a predetermined tile size, using instructions operating on fixed size tiles~\cite{libxsmm_ipdps20}.
The irregular nature of unstructured sparsity means we cannot know tile sizes a priori, and further, they may all be different. In the context of how software is written, this makes it tricky to define an easy-to-use ISA.

\begin{figure}[!t]
    \centering
    \includegraphics[width=0.45\textwidth]{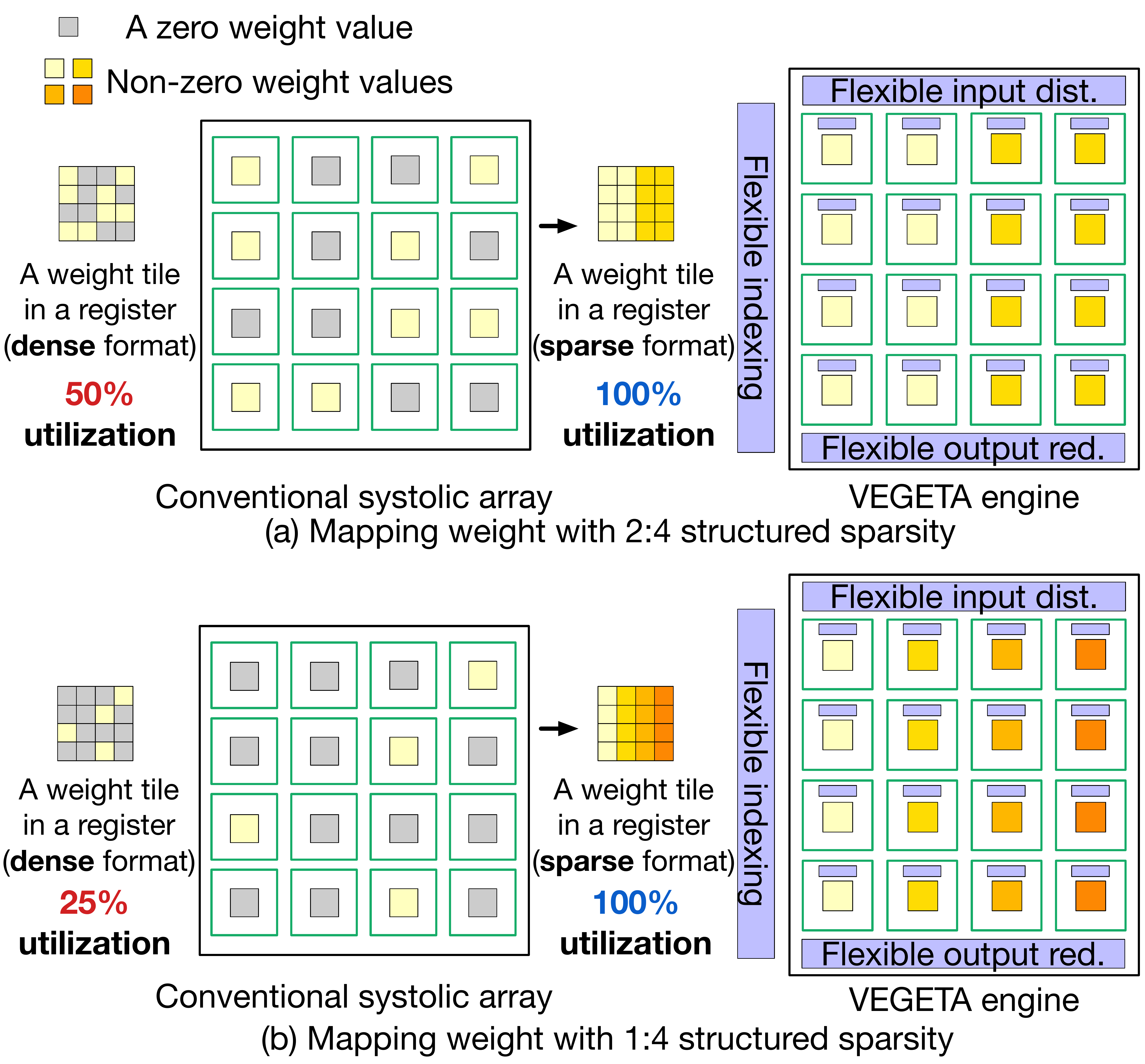}
    \vspace{-0.5em}
    \caption{Comparison of utilization of PEs in a Weight Stationary (WS) systolic array and VEGETA engines with sparse weights. 
    }
    \label{fig:underutlization-systolic}
    \vspace{-1.5em}
\end{figure}
\textbf{Challenge 2: Implementation overheads.} 
There are trade-offs between different options for the register file (RF) and systolic array (SA) when supporting sparsity.
The RF feeding the SA could hold a dense (i.e., conventional, with zeros) or sparse/compressed representation of each tile; if sparse, we need indexing logic and metadata to match each non-zero to the appropriate values from the other matrix.
Also, the SA could be comprised of conventional PEs (i.e., dense) or be enhanced to be sparsity-aware and skip zeros, at the cost of additional interconnects inside each PE~\cite{sigma}. Naturally, all these structures and interconnects add overhead.

To address these challenges, we make the following design decisions: (i) We limit our scope to sparsity support for DL workloads, where the typical sparsity degree is up to 95\%. 
Further, we add HW support for \textit{flexible $N$:$M$ structured sparsity}, leveraging insights from a recent work~\cite{domino2021nips} which has shown that adopting \textit{layer-wise} $N$:$M$ sparsity shows better accuracy compared to network-wise.
We also show how the HW can support unstructured sparsity by transforming the target matrix using row-wise $N$:$M$ sparsity.
(ii) We add sparsity support in \textit{both} the RF (\autoref{sec:isa}) and SA (\autoref{sec:architecture}) to achieve efficient utilization of the register storage and MACs.




\newcommand{\xmark}{\ding{55}}%
\newcommand{\cmark}{\ding{51}}%

\begin{table}[t] 
\scriptsize
\centering
\caption{Comparison of Sparsity Granularity of Previous works.}
\begin{tabular}{|c|c|c|c|c|}
\hline
\textbf{} 
&
\textbf{Network-wise}
&
\textbf{Layer-wise}
& 
\textbf{Tile-wise}
& 
\textbf{Row-wise}
\\
\hline
\hline

\begin{tabular}[c]{@{}c@{}}
NVIDIA STC~\cite{mishra2021accelerating}
\end{tabular}

& \cmark
& \xmark
& \xmark
& \xmark
\\
\hline

\begin{tabular}[c]{@{}c@{}}
STA~\cite{sta_cal20}
\end{tabular}
& \cmark
& \xmark
& \xmark
& \xmark
\\
\hline

\begin{tabular}[c]{@{}c@{}}
S2TA~\cite{s2ta_hpca22}
\end{tabular}
& \cmark
& \cmark
& \hspace{3pt}\cmark\tablefootnote{They do not claim they support tile-wise, but it can be extended.}
& \xmark
\\
\hline

\begin{tabular}[c]{@{}c@{}}
\textbf{VEGETA}
\end{tabular}
& \cmark
& \cmark
& \cmark
& \cmark
\\
\hline

\end{tabular}
\label{tab1}
\label{table:nm-hw}
\vspace{-1.5em}
\end{table}

\subsection{HW Support for Flexible $N$:$M$ Structured Sparsity} 
\label{subsec:hw_support}
\autoref{fig:underutlization-systolic} shows under-utilization challenges for a conventional WS systolic array when used with sparse weights.
A WS systolic array keeps the weight values stationary over time while inputs and outputs are streamed in a skewed manner. 
If sparse weights are used with 2:4/1:4 structured sparsity, 50\%/25\% of PEs are mapped with zero weights, which causes useless computation.
This work proposes an enhanced systolic array-based matrix engine which maps only non-zero weight values and distributes/reduces the correct input/output elements at the right time, leveraging the strengths of a systolic array, in the presence of some irregularity in the inputs.
This also requires logic in a PE to pick the right input elements for MACs.
The indexing logic and input distribution and output reduction logic to support flexible $N$:$M$ structured sparsity (i.e., the purple boxes in \autoref{fig:underutlization-systolic}) are presented in \autoref{sec:architecture}. 
In terms of sparsity granularity, we not only support network/layer/tile-wise, but also row-wise $N$:$M$ sparsity.
In \autoref{table:nm-hw}, we compare the supported sparsity granularity of our design and previous works that support $N$:$M$ sparsity.



\subsection{Transforming Unstructured to Row-Wise $N$:$M$ Sparsity} 
\label{subsec:transform}
While native support for unstructured sparsity can provide higher accuracy with extreme sparsity, the area overhead to support the sparsity in the form of a highly reconfigurable networks on chips (NoC)~\cite{sigma} and a sizable sparse controller \cite{extensor, geng2019awb} 
is only justifiable on standalone accelerators~\cite{sambanova, cerebras}, not within CPUs.
However, given an unstructured sparse tile, one can derive a row-wise $N$:$M$ sparse tile that covers all non-zeros in the given sparse tile by selecting appropriate $N$:$M$ per each row. 
For example, assuming 1:4, 2:4, and 4:4 are available sparsity patterns, one can analyze each row of the target unstructured tile to find the most sparse $N$:$M$ sparsity that covers all non-zeros in the row.
Then, each row can be compressed using the corresponding $N$:$M$ sparsity.
For example, the first and the second rows of \autoref{fig:sparsity} (a) would be compressed with 2:4 while the third and the fourth would be compressed with 1:4 to guarantee that none of non-zero values get lost.
This transformation does not cause any accuracy drop since it is lossless, meaning that all non-zeros in the original unstructured sparse matrix will still exist in the corresponding structured sparse matrix.
\autoref{fig:sparsity} (c) is derived using this transformation from \autoref{fig:sparsity} (a), covering all non-zeros (similarly, tile-wise 2:4 is used to derive \autoref{fig:sparsity} (b) from \autoref{fig:sparsity} (a)).
We use this transformation to leverage unstructured sparsity using VEGETA and show the estimated performance gain in \autoref{sec:row-wise-evalution}.

\subsection{VEGETA Design Overview}
This work presents VEGETA, 
which includes ISA extensions and microarchitecture support for structured/unstructured sparsity using flexible $N$:$M$ fine-grained structured sparsity HW~\cite{domino2021nips, zhou2021learning} in CPUs. We present a detailed design for some specific and important points (1:4, 2:4, 4:4) to explain detailed extensions for both ISA and the microarchitecture, but both can naturally be extended for different block sizes (\textit{M}).


We use a 32$\times$16 conventional WS systolic array as the baseline, inspired by RASA~\cite{rasa_dac21} and Intel's TMUL~\cite{intel2020isa}.
Comparing against a dense matrix engine rather than a vector engine provides a strong baseline due to the huge gap in compute throughput between typical matrix engines and vector engines (\autoref{sec:vector_matrix}).  
Our proposed VEGETA engine maintains the same number of MAC units as the baseline, adding the ability to skip zero-valued weight values via new control logic, multiplexers, and adders for reductions, along with some wider data paths.  
We target mixed-precision with BF16/FP32 which is widely used for both inference and training on commercial devices~\cite{intel2020isa, tpu-bf16, nvidia-bf16}.

\section{VEGETA Instruction Set Architecture}
\label{sec:isa}

\subsection{Register File Support}
\label{sec:rf}
\begin{figure}[!t]
    \centering
    \includegraphics[width=0.47\textwidth]{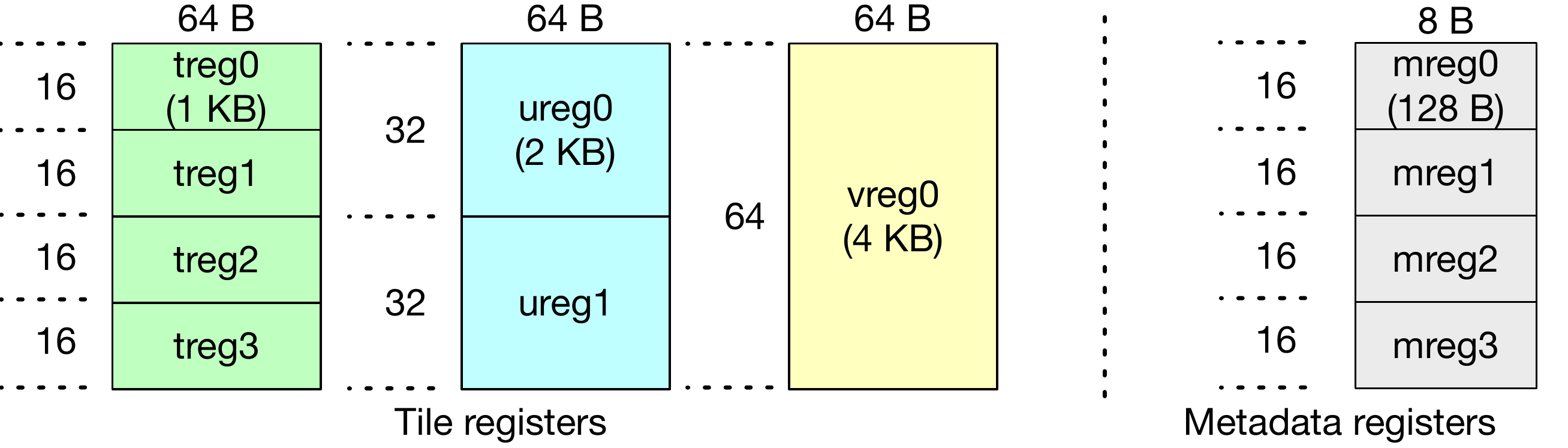}
    \vspace{-0.5em}
    \caption{VEGETA tile registers and metadata registers.}
    \label{fig:regs}
    \vspace{-1em}
\end{figure}
Inspired by Intel AMX~\cite{intel2020isa}, we assume there are eight 1 KB tile registers (treg0-7), each comprising 16 rows of 64 Bytes. 
We define a \textit{tile} as a 2D block comprising rows of data separated by a fixed stride. We define an \textit{effective tile} as the larger sparse tile captured by the non-zeros present in the compressed tile in the case of a sparse matrix (along with metadata). 
A tile register can hold 16$\times$32 BF16 elements or 16$\times$16 FP32 elements. To support 2:4 and 1:4 sparsity, we introduce aliased tile registers of size 2 KB (utile register or ureg) and 4 KB (vtile register or vreg), respectively. 
One ureg is composed of two consecutive tregs, while one vreg is composed of two consecutive uregs as shown in \autoref{fig:regs}.

Next, we introduce metadata registers (mreg0-7) to store metadata information for sparse tiles. As shown in \autoref{fig:compressed-matrix}, a pair of bits in the metadata represents the position of one non-zero element in a block of the compressed sparse matrix. Since a single row of the tile register holds 32 non-zero BF16 elements, the corresponding metadata register row holds $32\times2$ bits of metadata. 
Hence, an mreg has 16 rows, each with 64 bits of metadata for a total of 128 Bytes.
Note that while a treg can hold 16$\times$32 BF16 elements,
a treg and mreg for 2:4 sparsity can be used to store data for the effective tile whose dimension is 16$\times$64.
Similarly, when mreg is used for 1:4 sparsity, a treg and mreg can represent an effective tile whose dimension is 16$\times$128.

 \begin{table}[t] 
\scriptsize
\centering
\caption{VEGETA Instructions (MD refers to Metadata of a tile).
}
\begin{tabular}{|c|c|c|}
\hline
\textbf{Instruction} 
& \textbf{Operands} & \textbf{Explanation} \\
\hline
\hline
\tileloadt 
& \begin{tabular}[c]{@{}c@{}}
dst: treg, 
src: ptr[TILE]
\end{tabular}
& \begin{tabular}[c]{@{}c@{}}
Load 1 KB from 
ptr[TILE] to treg.
\end{tabular}
\\
\hline

\tileloadu
& \begin{tabular}[c]{@{}c@{}}
dst: ureg, 
src: ptr[TILE]
\end{tabular}
& \begin{tabular}[c]{@{}c@{}}
Load 2 KB from 
ptr[TILE] to ureg.  \\ 
\end{tabular}
\\
\hline

\tileloadv
& \begin{tabular}[c]{@{}c@{}}
dst: vreg, 
src: ptr[TILE]
\end{tabular}
& \begin{tabular}[c]{@{}c@{}}
Load 4 KB from 
ptr[TILE] to vreg.  \\ 
\end{tabular}
\\
\hline

\metaload
& \begin{tabular}[c]{@{}c@{}}
dst: mreg, 
src: ptr[MD]
\end{tabular}
& \begin{tabular}[c]{@{}c@{}}
Load 128 B from 
ptr[MD] to mreg.
\end{tabular}
\\
\hline

\tilestoret 
& dst: ptr[TILE], src: treg
& \begin{tabular}[c]{@{}c@{}}
Store 1 KB from 
ptr[TILE] to treg.\\ 
\end{tabular}
\\
\hline

\tilemm 
&
\begin{tabular}
[c]{@{}c@{}}dst, src0: treg,\\ src1: treg, src2: treg
\end{tabular}
&
\begin{tabular}[c]{@{}c@{}}
Multiply dense tile src1 \\
with dense tile src2, and add\\
the result back to tile dst.
\end{tabular}
\\
\hline

\tilespmmu 
& 
\begin{tabular}
[c]{@{}c@{}}dst, src0: treg,\\ src1: treg, src2: ureg
\end{tabular}
& \begin{tabular}[c]{@{}c@{}}
Multiply sparse tile src1 (2:4) \\
with dense tile src2, and add\\
the result back to tile dst.
\end{tabular}
\\
\hline

\tilespmmv
& 
\begin{tabular}
[c]{@{}c@{}}dst, src0: treg,\\ src1: treg, src2: vreg
\end{tabular}
& \begin{tabular}[c]{@{}c@{}}
Multiply sparse tile src1 (1:4) \\
with dense tile src2, and add\\
the result back to tile dst.
\end{tabular}
\\
\hline

\tilespmmr
& 
\begin{tabular}
[c]{@{}c@{}}dst, src0: ureg,\\ src1: treg, src2: ureg
\end{tabular}
& \begin{tabular}[c]{@{}c@{}}
Multiply sparse tile src1 (row-wise\\
$N$:4) with dense tile src2, and\\
add the result back to tile dst.
\end{tabular}
\\
\hline

\end{tabular}
\label{tab1}
\label{table:amx_instructions}
\vspace{-2em}
\end{table}

\subsection{VEGETA Instructions}
\label{sec:isa_extensions}
\autoref{table:amx_instructions} summarizes the VEGETA instructions using the aforementioned registers. 
\tileloadt, \tileloadu, and \tileloadv load a 1 KB, 2 KB, and 4 KB tile from the specified address to a treg, ureg, and vreg, respectively, while \tilestore stores a 1 KB tile from a treg to memory. 
\tileloadt can be used to load either a dense tile or the non-zero values of a compressed sparse tile. In the latter case, the 1 KB tile has an effective tile size of 2 KB or 4 KB for 1:4 and 2:4 sparsity ratios, respectively, as mentioned above. Furthermore, the load of a sparse tile must be accompanied by a corresponding \metaload instruction, which loads 128 B of metadata from memory to an mreg.


The \tilemm, \tilespmmu, and \tilespmmv instructions perform a tile matrix multiply and accumulate, $\boldsymbol{C}\mathrel{{+}{=}}\boldsymbol{A} \times \boldsymbol{B}$, where $\boldsymbol{A}$ and $\boldsymbol{B}$ are BF16 tiles and $\boldsymbol{C}$ is the FP32 output. 
$\boldsymbol{A}$ holds a tile of a sparse matrix.  With data in compressed format, $\boldsymbol{A}$ holds a fixed number of non-zeros, but with higher sparsity (smaller $N$:$M$), its {\it effective} size is larger.
In contrast, the actual size of the dense $\boldsymbol{B}$ tile must grow as $\boldsymbol{A}$ gets sparser. 
For example, assuming a dense $\boldsymbol{A}$ tile is $16\times32$, the effective size of $\boldsymbol{A}$ for a 2:4 sparsity ratio is $16\times64$ (2 KB), while the non-zero values could still fit into a 1 KB treg.
Thus, the corresponding $\boldsymbol{B}$ tile should be $64\times16$, which will fit into a 2 KB ureg.
Similarly, for a 1:4 ratio, the effective size of $\boldsymbol{A}$ is 4 KB and the corresponding $\boldsymbol{B}$ tile will fit into a 4 KB vreg. 
Note that the corresponding output tile $\boldsymbol{C}$ is a constant size ($16\times16, FP32$) and fits in a 1 KB treg.
We call \tilemm a VEGETA tile GEMM instruction and \tilespmmu and \tilespmmv VEGETA tile SPMM instructions.


To summarize, \tilemm performs a dense (4:4) GEMM operation on 1 KB treg inputs, while \tilespmmu performs an SPMM operation where $\boldsymbol{A}$ is a 2:4 compressed sparse 1 KB tile, $\boldsymbol{B}$ is a dense 2 KB tile, and the output $\boldsymbol{C}$ is a dense 1 KB tile.
Thus, \tilespmmu calculates $\boldsymbol{C}$ $(16 \times 16, FP32)\mathrel{{+}{=}}\boldsymbol{A}$ $(16 \times 64, BF16) \times \boldsymbol{B}$ $(64 \times 16, BF16)$.
Similarly, \tilespmmv calculates $\boldsymbol{C}$ $(16 \times 16, FP32)\mathrel{{+}{=}}\boldsymbol{A}$ $(16 \times 128, BF16) \times \boldsymbol{B}$ $(128 \times 16, BF16)$.
Note that $\boldsymbol{C}$ is used as both input and output.

\begin{minipage}[t]{\linewidth}
\begin{center}
  \begin{lstlisting}[caption={An implementation of SPMM.}\label{lst:spmm_base},escapechar=\%]
    for {int i = 0; i < $\text{D}_\text{m}$/$\text{T}_\text{m}$; i++} {
      for {int j = 0; j < $\text{D}_\text{n}$/$\text{T}_\text{n}$; j++} {
        for {int k = 0;k < $\text{D}_\text{k}$/$\text{T}_\text{k}$; k++} {
          // C[i][j] += A[i][k] + B[k][j]
          %\tileloadu% ureg0, tileB[j][k];
          %\tileloadt% treg2, tileC[i][j];
          %\tileloadt% treg3, tileA[i][k];
          %\metaload% mreg3, metadataA[i][k];
          %\tilespmmu% treg2, treg3, ureg0;
          %\tilestoret% tileC[i][j], treg2;
        }
      }
    }
  \end{lstlisting} 
\end{center}
\end{minipage}

The number of useful MAC operations required to calculate $\boldsymbol{C}$ is the same for \tilemm, \tilespmmu, and \tileloadv (8192).
For each output element, the number of effectual MAC computations is 32.
Finally, to support SPMM with a row-wise $N$:$M$ sparse matrix $A$, we introduce \tilespmmr. \tilespmmr calculates $\boldsymbol{C}$ $(R \times 16, FP32)\mathrel{{+}{=}}\boldsymbol{A}$ $(R \times 64, BF16) \times \boldsymbol{B}$ $(64 \times 16, BF16)$ where $R$ can vary from 8 to 32, depending on $N$:4 sparsity for each row (which will be stored as extra metadata, 32$\times$2 bits, or 8B, at most).

We show an example SPMM kernel assuming $\boldsymbol{A}$ with 2:4 sparsity using VEGETA instructions in \autoref{lst:spmm_base}. 
$D_m$, $D_n$, and $D_k$ indicate the size of each dimension while $T_m$, $T_n$, and $T_k$ show the corresponding tile sizes. 
In this case, $T_m$, $T_n$, and $T_k$ are 16, 16, and 64, respectively.
We store the values of matrix $B$ in a transposed manner in the tile registers.
We implemented an optimized version of this kernel to evaluate our architecture that we introduce in \autoref{sec:architecture}.



\subsection{Flexibility in the Block Size, M}
In this work, we assume $M=4$ to explain our extension in detail, which can handle different fine-grained structured sparsity patterns including 1:4/2:4/4:4, but our extension is not limited to the specific size of $M$.
While we use $M=4$ in our implementation, our approach can be extended to $M=2^m$ by modifying the tile/metadata registers and ISA to support larger input registers. A larger $M$ provides greater flexibility to the sparse model design and may result in 
improved accuracy~\cite{domino2021nips}, but would cost more HW.

\section{VEGETA Engine Architecture}
\label{sec:architecture}

\subsection{Processing Units and Processing Elements} \label{subsec:pu-pe}
\begin{figure*}[!t]
    \centering
    \includegraphics[width=0.95\textwidth]{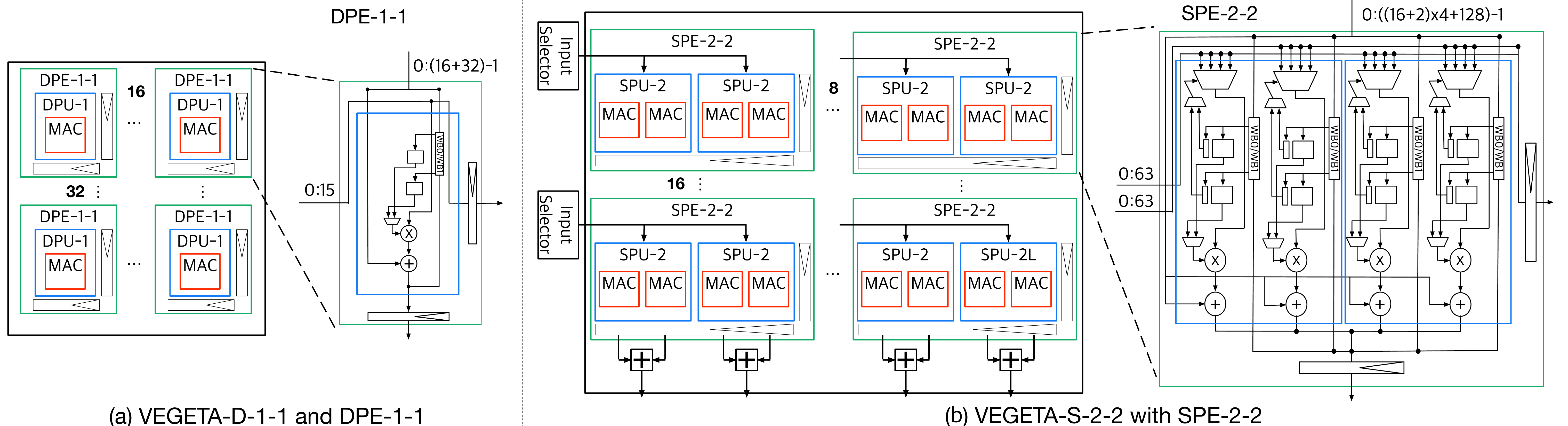}
    \caption{Two VEGETA designs: VEGETA-D-1-1 and VEGETA-S-2-2.}
    \label{fig:vegeta-designs}
\end{figure*}
\begin{table*}[t] 
\scriptsize
\centering
\caption{Different VEGETA-D and VEGETA-S Designs.}
\begin{tabular}{|c|c|c|c|c|c|c|c|c|c} 
\hline
\textbf{VEGETA Engine} 
& \textbf{$N_{rows}$}
& \textbf{$N_{cols}$}
& \begin{tabular}[c]{@{}c@{}}
\textbf{\# of MACs per PE} \\
\textbf{($\alpha\times\beta$)}
\end{tabular}
& \begin{tabular}[c]{@{}c@{}}
\textbf{\# of inputs} \\
\textbf{per PE}
\end{tabular}
& \begin{tabular}[c]{@{}c@{}}
\textbf{Broadcast} \\
\textbf{Factor ($\alpha$)}
\end{tabular} & 
\begin{tabular}[c]{@{}c@{}}
\textbf{Drain} \\
\textbf{Latency}
\end{tabular} & 
\begin{tabular}[c]{@{}c@{}}
\textbf{Supported} \\
\textbf{Sparsity}
\end{tabular} & 
\textbf{Example from Prior Work} \\
\hline
\hline
VEGETA-D-1-1
& 32
& 16
& 1
& 1
& 1
& 16
& Dense
& \begin{tabular}[c]{@{}c@{}}
Conventional SA~\cite{systolic},
RASA-SM~\cite{rasa_dac21}
\end{tabular}
\\
\hline

VEGETA-D-1-2
& 16
& 16
& 2
& 2
& 1
& 16
& Dense
& \begin{tabular}[c]{@{}c@{}}
RASA-DM~\cite{rasa_dac21}
\end{tabular}
\\
\hline

VEGETA-D-16-1
& 32
& 1
& 16
& 1
& 16
& 1
& Dense
& \begin{tabular}[c]{@{}c@{}}
Intel TMUL~\cite{intel2020isa}-Inspired Unit
\end{tabular}
\\
\hline

VEGETA-S-1-2
& 16
& 16
& 2
& 8
& 1
& 16
& 1:4, 2:4, 4:4
& New design

\\ 
\hline

VEGETA-S-2-2
& 16
& 8
& 4
& 8
& 2
& 8
& 1:4, 2:4, 4:4
&  New design
\\ 
\hline

VEGETA-S-4-2
& 16
& 4
& 8
& 8
& 4
& 4
& 1:4, 2:4, 4:4
& New design
\\ 
\hline

VEGETA-S-8-2
& 16
& 2
& 16
& 8
& 8
& 2
& 1:4, 2:4, 4:4
& New design
\\
\hline

VEGETA-S-16-2
& 16
& 1
& 32
& 8
& 16
& 2
& 1:4, 2:4, 4:4
& New design
\\
\hline

\end{tabular}
\vspace{-1em}
\label{tab1}
\label{table:vegeta_designs}
\end{table*}
\label{subsec:pe}
\textbf{Processing Unit (PU).} A PU is composed of a number of MAC units that contribute to the same output element.
In a WS systolic array, the dot product to produce each output element is mapped to a column of PUs. Partially accumulated results trickle down a column, and the final result of the dot product exits the bottom of the array.
In a conventional design, a PU is composed of one MAC unit, the height of the array is the length of the dot product, and each PU produces a single partial sum each cycle.
We can, however, break the set of operations for each dot product into pieces, or ``lanes.'' If we pack multiple MAC units into a PU, each PU can work on all lanes in parallel, and we can scale down the height of the systolic array.
In this case, we must combine the partial sum for each lane at the bottom of the systolic array.
We call the number of lanes or the number of MAC units in a PU, or the \textit{reduction factor}, $\beta$.
$\beta$ also indicates how many partial sums need to be reduced at the bottom of the systolic array to generate a single output element.

\textbf{Processing Element (PE).} 
We group PUs that share the same eastbound inputs and output buffers into a PE.
That is, peer-to-peer forwarding happens between PEs, and the input fed from the west is broadcasted to all PUs in a PE.
The more PUs in a PE, the narrower the systolic array, and the more we amortize the overhead of the horizontal pipeline buffer.
This improvement in area and power comes at the cost of lower achievable frequency since the broadcast must reach more PUs.
We call the number of PUs in a PE, or the \textit{broadcast factor}, $\alpha$.
We label PE designs as PE-$\alpha$-$\beta$. 
For example, PE-1-1 indicates each PE has one single-MAC PU, while PE-4-2 indicates each PE has four two-MAC PUs (PU-2s).

\textbf{SPU and SPE.}
We enhance PUs and PEs to support tile SPMM with flexible $N$:$M$ structured sparsity. 
To distinguish them, we call a PU and PE without sparsity support Dense Processing Unit (DPU) and Dense Processing Element (DPE), respectively, while we call a PU and PE with sparsity support Sparse Processing Unit (SPU) and Sparse Processing Element (SPE), respectively.
DPEs and SPEs are the building blocks for VEGETA-D (for dense) and VEGETA-S (for dense and sparse) engines which we explain in the following sub-sections.
The main differences between a DPE and SPE are $M$ to 1 MUXes and a metadata buffer added to each weight buffer.
Each cycle, an SPE receives multiple input elements, and uses this extra hardware to select, for each weight, the one corresponding to its index.
To support flexible $N$:$M$ structured sparsity, we choose $\beta$ as $\frac{M}{2}$; this ensures that input elements need only be fed into a single row.
Since we use $M=4$, we use $\beta=2$ for our SPEs.
In summary, we use DPE-$\alpha$-$\beta$ and SPE-$\alpha$-$\beta$ to indicate the broadcast factor ($\alpha$) and the reduction factor ($\beta$) of each PE.
For example, the broadcast factor ($\alpha$) of SPE-2-2 (shown in ~\autoref{fig:vegeta-designs} (b)) is 2, which indicates that a single SPE-2-2 is composed of two ($=\alpha$) SPU-2s.
\begin{figure*}[t]
    \centering
    \includegraphics[width=0.99\textwidth]{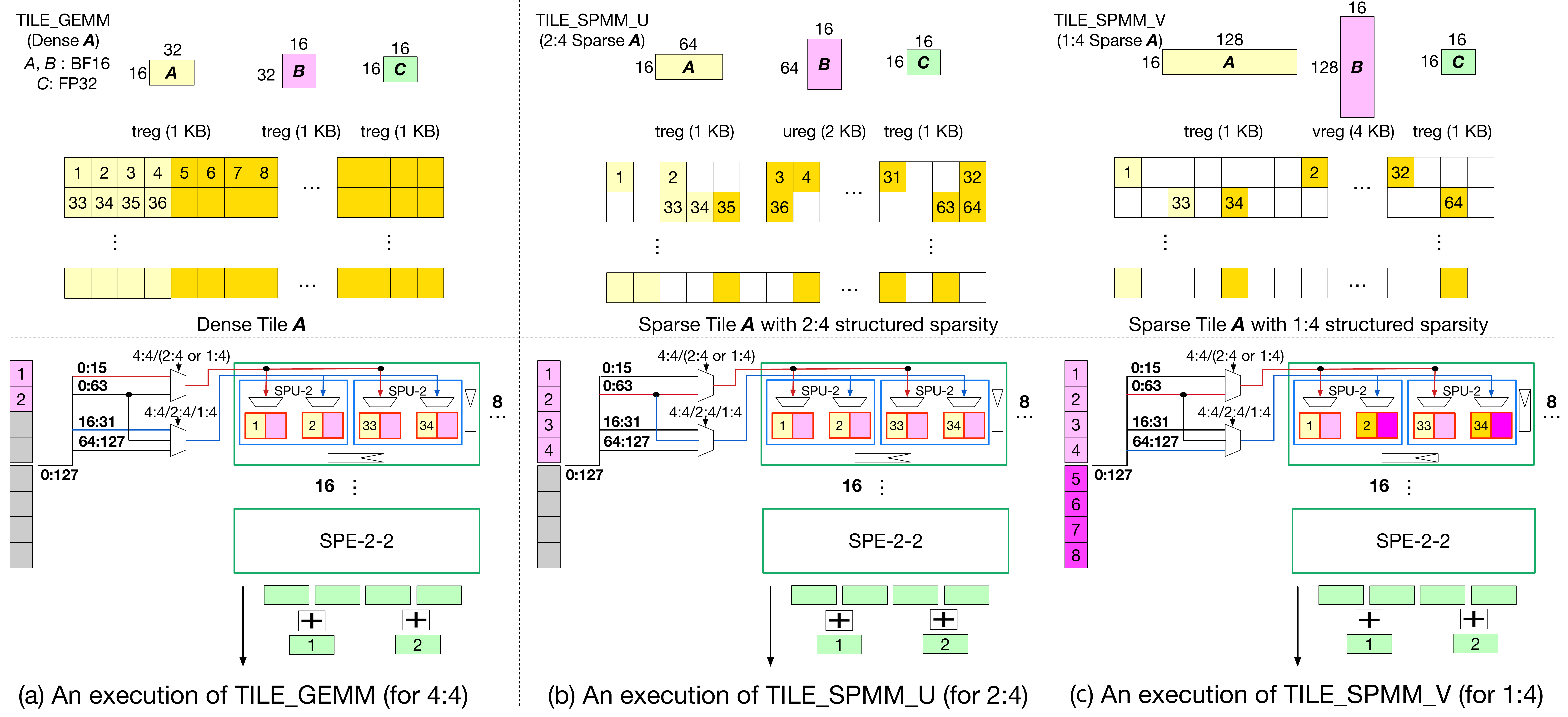}
    \vspace{-1em}
    \caption{Comparison of executions of \tilemm, \tilespmmu, and \tilespmmv on VEGETA-S-2-2.}
    \label{fig:input-selection}
    \vspace{-1em}
\end{figure*}
\textbf{Execution flow of DPE/SPE}:
DPE-1-1, a conventional single-MAC PE, operates as follows.
First, a weight element is loaded into its weight buffer. 
Then, an input element and partial sum element are streamed from the left and top ports, respectively.
The input element is multiplied with the element in the weight buffer and accumulated with the partial sum element.
The generated partial sum flows to the PE below the current PE, and the input element flows to the PE on the right.

An SPE is composed of $\alpha$ SPUs, each of which comprises $\beta$ MAC units.
Different SPUs calculate partial sums in parallel for the \textit{different} output elements, while each MAC unit in an SPU calculates partial sums in parallel for the \textit{same} output element.
An SPE-2-2 receives two ($=\beta$) input \textbf{blocks} (instead of one input \textbf{element}) from the west and broadcasts them to $\alpha$ SPUs.
An input block is fed into each $M$ to 1 MUX in an SPU, and using the metadata for the corresponding weight value in the weight buffer, the corresponding input element from the block is selected. 
Then, each selected input element is multiplied by the corresponding weight element and accumulated.
Since each SPU generates $\beta$ partial sums, and there are $\alpha$ SPUs, it forwards a total of $\alpha\times\beta$ partial sums (for example, SPE-2-2 will generate $2\times2=4$ partial sums) to the south port.
\subsection{VEGETA Matrix Engine for Tile-Wise $N$:$M$ Sparsity}

\begin{figure*}[!t]
    \centering
    \includegraphics[width=0.99\textwidth]{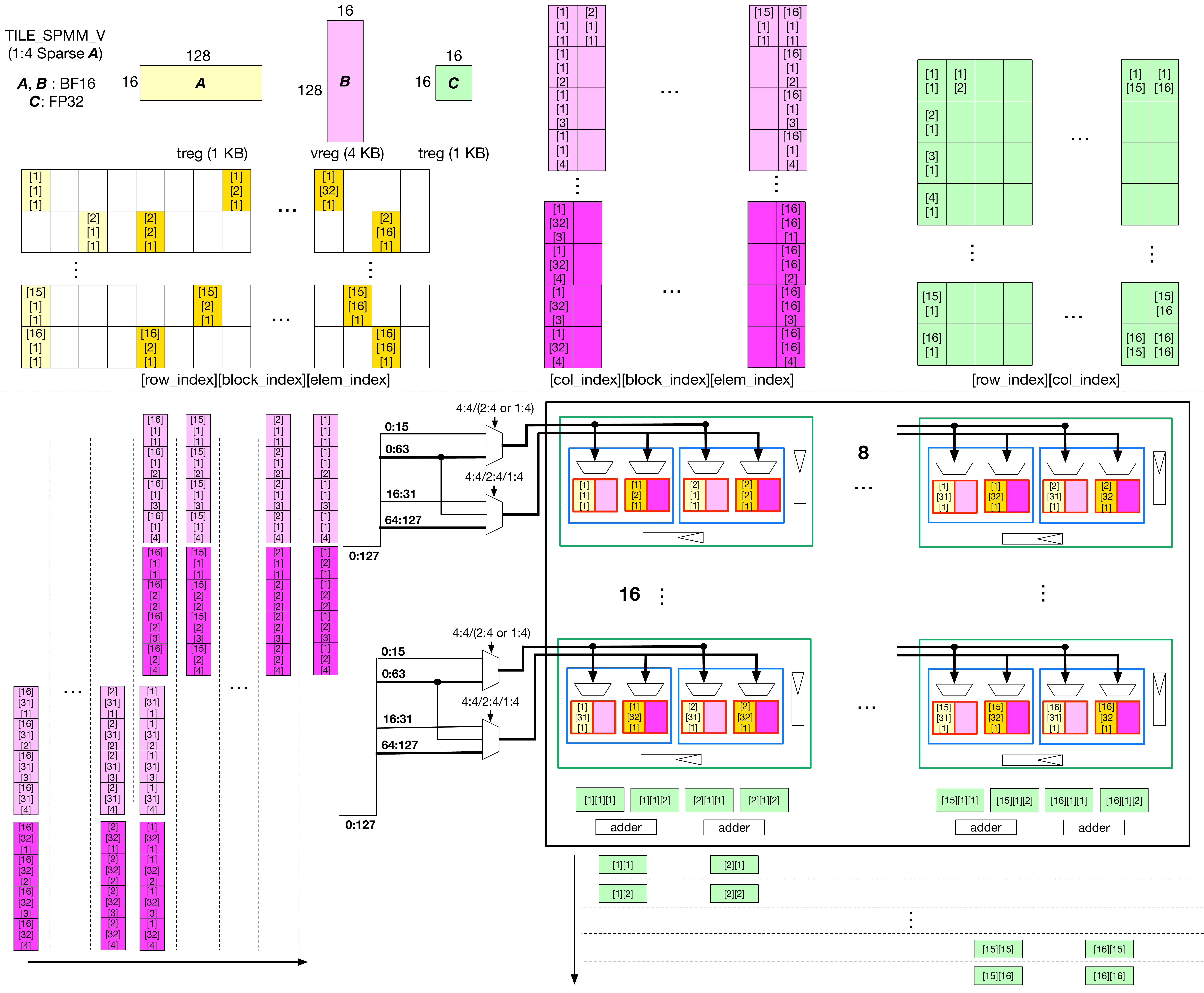}
    \caption{Cycle-level visualization for  \tilespmmv instruction on VEGETA-S-2-2 with 1:4 structured sparsity for matrix $\boldsymbol{A}$ with dimensions $\boldsymbol{A}$: 16$\times$128 (yellow), $\boldsymbol{B}$: 128$\times$16 (magenta), $\boldsymbol{C}$:16$\times$16 (green). 
    }
    \vspace{-1em}
    \label{fig:spmm-cycles}
\end{figure*}
Using DPEs and SPEs, we show how to build various VEGETA engines.
We use -\{S (for sparse) $|$ D (for dense)\}-$\alpha$-$\beta$ to indicate the type of the base PE of a VEGETA engine.
For example, VEGETA-D-1-1 has DPE-1-1s as its PEs while VEGETA-S-2-2 is composed of SPE-2-2s.
In~\autoref{fig:vegeta-designs}, we show VEGETA-D-1-1 and VEGETA-S-2-2 as examples.
Note that adders (or adder trees) are needed at the bottom of the VEGETA engine if the reduction factor $\beta > 1$ to generate the final output element by reducing partial sums.
A VEGETA engine is a 2D array of $N_{rows} \times N_{cols}$ PEs, where neighboring PEs are connected.
Since a column of SPUs cooperates to calculate a single output element, $N_{rows}$ can be derived as $
    N_{rows} = \frac{\text{\# of effectual computations per output element}}{\beta}$.
The number of effectual computations per output element is 32 for the VEGETA Tile GEMM/SPMM instructions.
With $N_{rows}$, the number of PUs in a PE ($\alpha$), the number of MACs in a PU ($\beta$), and the total number of MAC units in a VEGETA engine, $N_{cols}$ can be derived 
    $N_{cols} = \frac{\text{\# of total MAC units}}{N_{rows} \times \alpha \times \beta}$.
In~\autoref{table:vegeta_designs}, we summarize different VEGETA engine design choices.

Each row of a VEGETA-S engine has a special unit called an Input Selector to select the right input blocks to support flexible structured sparsity.
In~\autoref{fig:input-selection}, we show how different VEGETA tile GEMM/SPMM instructions are executed on VEGETA-S-2-2 focusing on one SPE and the corresponding input selector.
For all three cases, non-zero values in each row of a weight tile are mapped onto a column of SPU-2s (i.e. 2 columns of MAC units). For example, weight elements 1-32 are mapped onto the first column of SPU-2 while weight elements 33-64 are mapped onto the second column of SPU-2.

\textbf{4:4 structured sparsity}: For \tilemm (4:4 sparsity), the weight tile is dense; thus, there are four effectual computations for an output element per input block (size of 4).
Thus, a half input block (two input elements) is fed into a row of PEs from the west port, and they are broadcast to each SPU in an SPE.
Since a weight tile is dense for \tilemm, the first element of an input block is multiplied with the first weight element in an SPU while the second element of the input block is calculated with the second weight element in the SPU.
Similar to a classic systolic array, the input is fed in a skewed manner so that the reduction for partial sums happens in a spatio-temporal manner along a column of MAC units. 
Once it reaches the bottom of the array, the partial sums from each MAC column in an SPU are accumulated in a reduction unit (an adder in this case) to get the final output element.

\textbf{2:4 structured sparsity}: For \tilespmmu (2:4 sparsity), an input block (instead of a half block) will be fed into a row of PEs from the left, and they are broadcast to each SPU in a SPE.
4 to 1 MUXes in an SPU choose the corresponding input element from an input block to be used in a MAC unit. 
For this case, the same block is used between MAC units in an SPU since there are two effectual computations for an output element per input block due to the 2:4 structured sparsity.

\textbf{1:4 structured sparsity}: Finally, for \tilespmmv (1:4 sparsity), there is only one effectual computation for an output element per input block.
Thus, two input blocks will be fed into a row of PEs from the left, and they are broadcast to each SPU in an SPE. 
Each MAC unit in an SPU gets an input block and chooses and computes with the corresponding input element in the block. 
Unlike the \tilespmmu, there is only one non-zero element in a weight block due to the 1:4 structured sparsity. 
Thus, two weight elements in an SPU are belonging to two different weight blocks, which implies that they need two different input blocks for computation.
In~\autoref{fig:spmm-cycles}, we show a detailed cycle-level visualization about how a VEGETA-S-2-2 executes a \tilespmmv, which computes tile SPMM with 1:4 sparse $\boldsymbol{A}$ (weight), dense $\boldsymbol{B}$ (input), and dense $\boldsymbol{C}$ (output).

\subsection{Optimizations}
\textbf{Pipelining.}
So far, we have shown how one VEGETA tile GEMM/SPMM instruction can be executed on a VEGETA engine.
For modest-sized tiles, filling and draining the systolic array for a tile GEMM/SPMM instruction can significantly lower PE utilization.
A recent work RASA~\cite{rasa_dac21} introduced pipelining the execution of tile GEMM instructions on a dense systolic array-based matrix engine; this overlaps different stages of execution for different instructions.
We extend the pipelining concept and show how different tile SPMM instructions can be executed concurrently.

Similar to the stages defined in the previous work, Weight Load (WL), Feed First (FF), Feed Second (FS), and Drain (DR) stages are used for pipelining both VEGETA-D designs and VEGETA-S designs.
WL is the first stage of the execution on the systolic array where weight elements (which will be stationary) are loaded to the corresponding PEs from north ports, which takes $N_{rows}$ cycles.
Next, during the FF stage, the corresponding inputs and output elements are getting fed from the west and north ports, which takes $T_n$ cycles, where $T_n$ is the number of columns in an input tile. 
This stage ends when the top-left PE stops receiving input/output elements.
Since the inputs are streamed in a skewed manner, the remaining rows of the systolic array are still receiving the new input elements from the west ports.
This stage is called as the FS stage and ends when there is no more input element coming in from the west ports, which takes $N_{rows}-1$ cycles.
Finally, to finish the remaining calculations in the systolic array, it needs $N_{cols}$ cycles during the DR stage followed by cycles required for reduction at the bottom.
Note that we need an additional stage for reduction after DR stage to wait for the remaining values in reduction units.
Also, unlike the conventional systolic array where each PE is receiving one input, weight, and output element at a time, $\beta$ input blocks, $\alpha$ weight elements, $\alpha$ output elements should be fed to each SPE.
~\autoref{fig:spmm-cycles} shows the steaming pattern of input element blocks.

$N_{rows}$ and $N_{cols}$ can be reduced by increasing $\alpha$ and $\beta$, larger $\alpha$ and $\beta$ might reduce latency of a single instruction. 
Larger $\alpha$ causes lower frequency while larger $\beta$ causes longer reduction latency and larger reduction unit, so we cannot arbitrarily increase those. 
Furthermore, to properly overlap different stages, it is crucial to balance the latencies of different stages.
Different VEGETA-S designs can use same pipelining stages since they follow the same streaming pattern. 

\begin{figure}[t]
    \centering
    \includegraphics[width=0.45\textwidth]{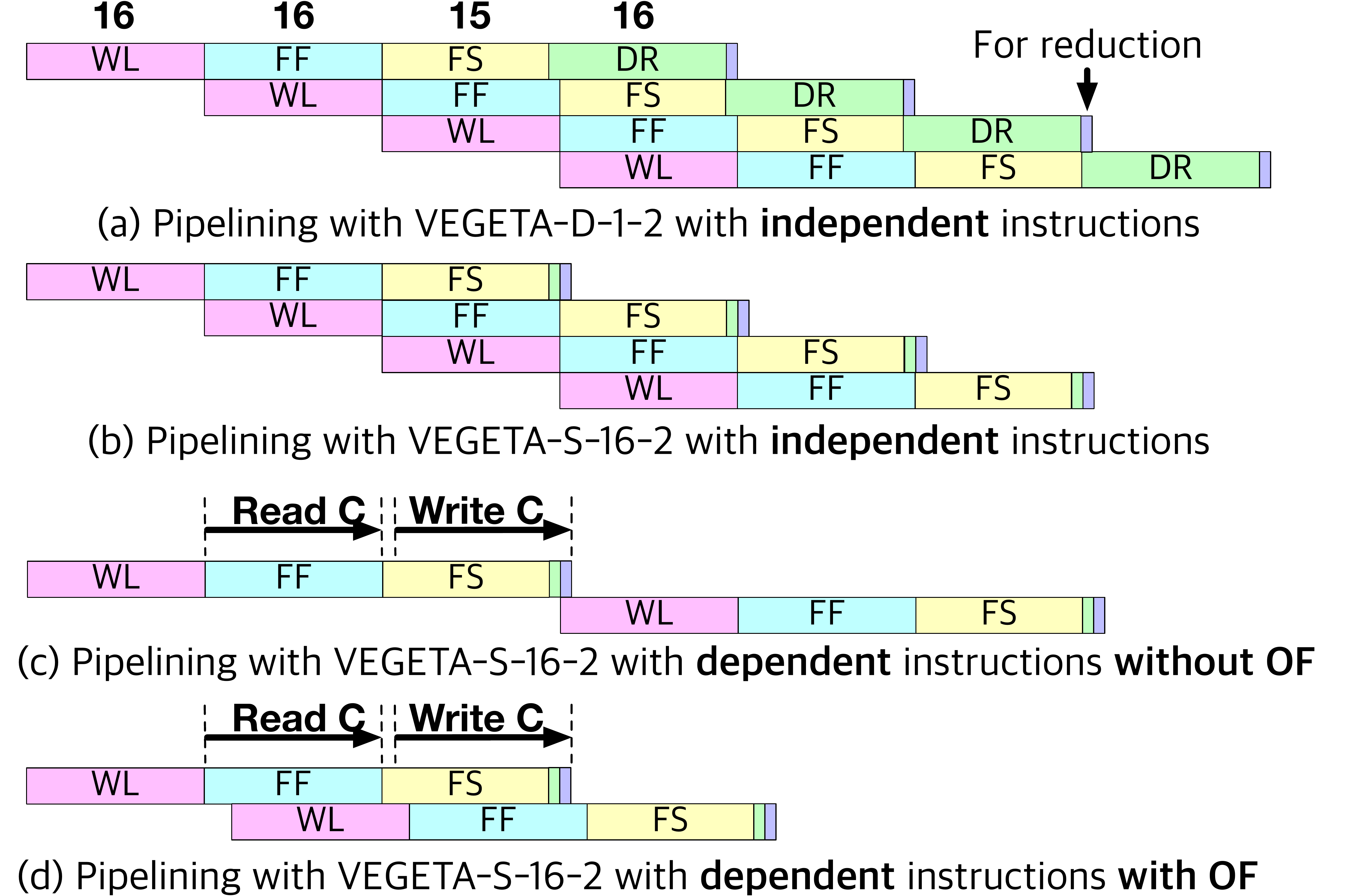}
    \caption{Pipelining on VEGETA-D-1-2/VEGETA-S-16-2.}
    \label{fig:pipelining}
    \vspace{-1em}
\end{figure}

~\autoref{fig:pipelining} (a) and (b) shows the pipelining examples for VEGETA-D-1-2 and VEGETA-S-16-2, assuming no dependency between instructions.
With pipelining, multiple instructions may be executed in the systolic array concurrently, but no two of them may be in the same stage of execution, since that would oversubscribe at least one hardware resource.  Since the latency of each stage is known, the scheduler can easily enforce this.
We observe that the next instruction can be executed after 16 cycles for VEGETA-S-16-2, which is same as VEGETA-D-1-2 since it is limited by the total number of MAC units (maximum compute throughput).
Due to the smaller $N_{rows}$ and $N_{cols}$, the latency of each instruction for VEGETA-S-16-2 is shorter than that of VEGETA-D-1-2.

\textbf{Output forwarding.}
Pipelining allows concurrent execution of independent instructions. 
However, it is often not allowed if the instructions have dependence between them.
Since VEGETA tile GEMM/SPMM instructions perform accumulation, the destination register is a source as well;
thus, for two back-to-back tile multiplication instructions with the same destination register, the second cannot begin execution until the first finishes.
A traditional approach to resolve this kind of pipeline stall is data forwarding.
We extend this ``output forwarding (OF)'' concept to matrix engines. 
In~\autoref{fig:pipelining} (c) and (d), we show the pipelining examples of dependent instructions for VEGETA-S-16-2 with and without output forwarding.
Since we feed a $\boldsymbol{C}$ tile into the systolic array over multiple cycles, we need to only be sure that particular $\boldsymbol{C}$ elements will be ready before we need to feed them.
A key insight for this is that the register reads/writes of $\boldsymbol{C}$ tile follow the exactly same order in a systolic array.
This is because every output element will be calculated $N_{rows}+log_2{\beta}$ cycles after it gets fed into the systolic array.
For example, when executing a VEGETA-S-16-2, a \tilespmmu starts reading the C tile when FF stage begins (Cycle $N_{rows}+1$). 
In the same order, the C tile will be written back from Cycle $2N_{rows}+log_2{\beta}$ so that the dependent instruction can start reading the same $\boldsymbol{C}$ tile.
This can significantly reduce the stalls which might have occurred between instructions with a dependency without OF, but a bypass buffer would be required to keep and forward the data before writing it back to the tile register.

\subsection{Flexibility in the Block Size, M}
Similar to the case for VEGETA ISA to support a broader range of sparsity,  
VEGETA engines can be easily extended for larger $M$. 
For example, to support the $M=16$ case (for 1:16, 2:16, 4:16, 8:16, and 16:16), 
a 16-to-1 MUX (or five 4-to-1 MUXes) is needed per MAC unit, which would take an input block composed of 16 elements and select the corresponding input element.
$\alpha$ and $\beta$ should be configured considering the balance of pipeline stages, data reuse, critical path, etc.

\subsection{Row-Wise $N$:$M$ Sparsity Support}
\label{subsec:row-wise-support}
\begin{figure}[!t]
    \centering
    \includegraphics[width=0.5\textwidth]{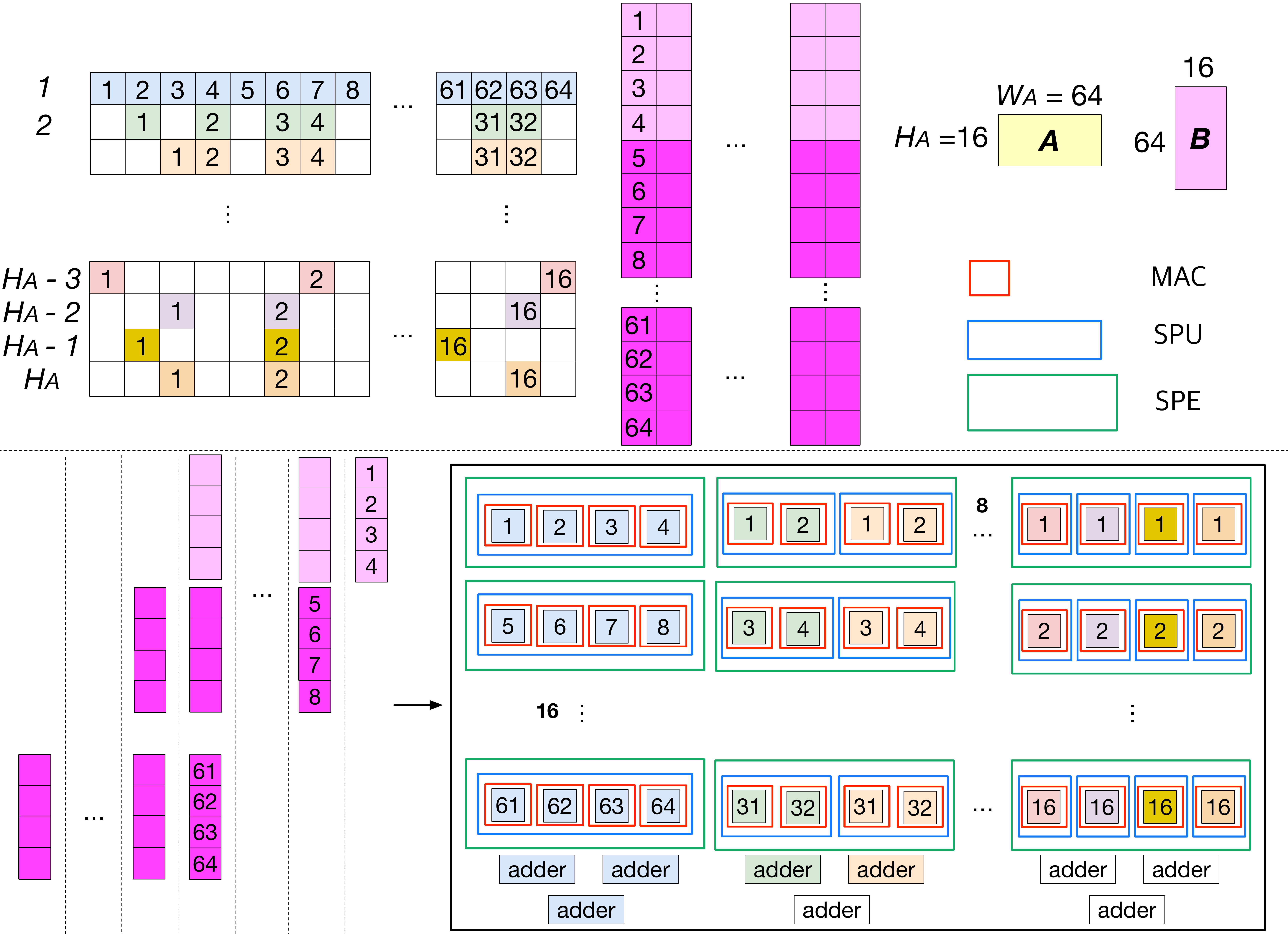}
    \caption{Mapping of row-wise $N$:$M$ sparse matrix on VEGETA-S.}
    \label{fig:Row-wise-vegeta-mapping}
    \vspace{-1em}
\end{figure}

VEGETA engines support row-wise $N$:$M$ sparsity, which can be used for leveraging unstructured sparsity using the method described in \autoref{subsec:transform}.
In \autoref{fig:Row-wise-vegeta-mapping}, we show a row-wise sparse matrix using 4:4, 2:4, 1:4 at Row 1, Row 2/3, and Row $H_{A}$-3 to $H_{A}$, respectively.
A row with 4:4 maps to an SPE-1-4 column, and outputs of the four SPU columns reduce to generate one partial sum. 
A row with 2:4 maps to an SPE-2-2 column, and outputs of each pair of SPU columns reduce, resulting in two partial sums per SPE column. 
A row with 1:4 maps to an SPE-4-1 column, generating 4 partial sums per SPE column. 
Using this mapping, we can ensure all columns are fully utilized while allowing different $N$:$M$ sparsity across different rows. 
The number of columns of a weight tile, $W_A$, is equal to $M\times N_{rows}$ to keep all columns fully utilized.
 
Similar to \autoref{fig:input-selection} (b), an input block (4 elements) is fed into a row of SPEs from the left and broadcast to each MAC in an SPE.
In \autoref{fig:Row-wise-vegeta-mapping}, the number of rows of a weight tile can vary based on the $N$:$M$ sparsity combinations in a tile.
When a row-wise $N$:$M$ sparse tile has $N_{4:4}$ rows with 4:4 sparsity, $N_{2:4}$ rows with 2:4 sparsity and $N_{1:4}$ rows with 1:4 sparsity, the number of columns in VEGETA-S and the number of rows of the weight tile, $H_{A}$ can be derived as $N_{cols} = N_{4:4} + \frac{N_{2:4}}{2} + \frac{N_{1:4}}{4}$ and $H_{A} = N_{4:4} + N_{2:4} + N_{1:4}$.
$H_A$ can vary from 8 to 32 depending on the sparsity degree of the tile.
Our approach requires having consecutive groups of rows of the weight tile with the same sparsity degree (e.g., 2 for 2:4 and 4 for 1:4); we call this as pseudo row-wise $N$:$M$ sparsity. 
We can employ a simple reordering in input/output DMA engines to group input rows with the same sparsity, and reorder the output elements back to their original order. Since this only needs an extra row of adders, the hardware overhead would be negligible compared to VEGETA-S which supports tile-wise flexible $N$:$M$.
\subsection{Integration of VEGETA Engines with CPU}
\begin{figure}[!t]
    \centering
    \includegraphics[width=0.48\textwidth]{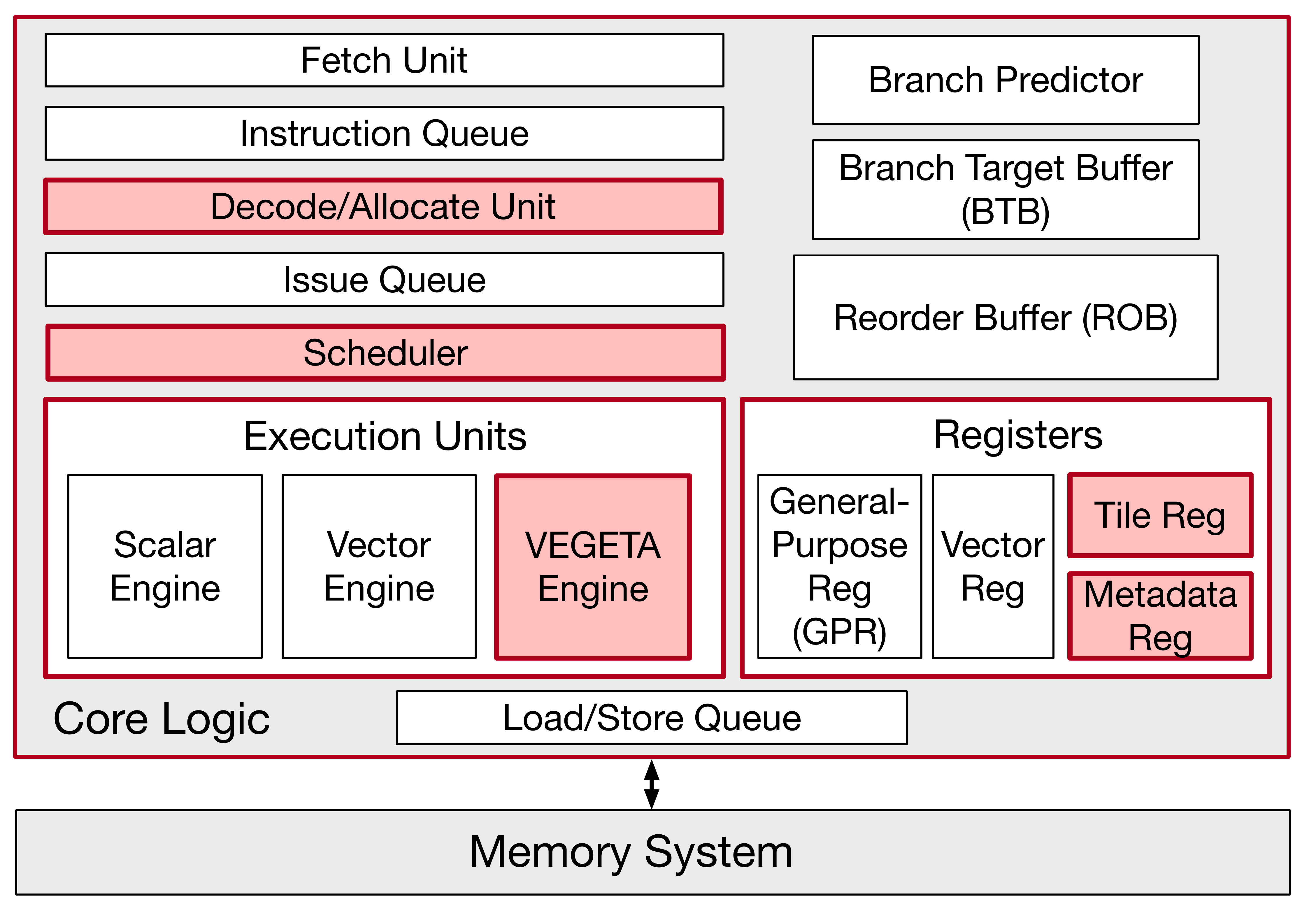}
    \caption{Overview of VEGETA in a CPU. We highlight the parts including our contributions with red.}
    \label{fig:vegeta-system-overview}
    \vspace{-1em}
\end{figure}
In~\autoref{fig:vegeta-system-overview}, we show how we integrate VEGETA in a CPU.
We also marked the components that we modify or introduce to integrate VEGETA in red.
Our baseline core design maintains separate physical and architectural register files. 
Also, we extend this to tile registers and metadata registers for VEGETA; this includes enhancements to the register file itself, allocator, renamer, scheduler, and ROB to add the ability to manage the new registers, track dependencies on them, and support precise exceptions.
We also enhance the scheduler to track static and dynamic information of instructions being executed in VEGETA engines to support pipelining and output forwarding at the right timing without interrupting scalar/vector instructions.
A \tileloadt (or \tilestoret) will be converted into 16 memory requests and each will be loading (or storing) 64 Bytes (cache line size) through load/store queue, not imposing any extra implication on cache/memory coherence/consistency.


\section{Evaluation} \label{sec:eval}
\begin{table}[t]
\scriptsize
\centering
\caption{Dimensions of DNN layers used in this evaluation.}
\vspace{-1em}
\begin{tabular}{|c|c|c|c|}
\hline
\textbf{Workloads} & \textbf{Dimensions} & \textbf{\# of MACs} \\
\hline
\hline
ResNet50-L1 & K=64, C=256, Y=56, X=56, R=1, S=1 & 51,380,224  \\
\hline
ResNet50-L2 & K=64, C=64, Y=56, X=56, R=3, S=3 & 115,605,504    \\
\hline
ResNet50-L3 & K=256, C=64, Y=56, X=56, R=1, S=1 & 51,380,224    \\
\hline
ResNet50-L4  & K=128, C=128, Y=28, X=28, R=3, S=3 & 115,605,504   \\
\hline
ResNet50-L5  & K=512, C=128, Y=28, X=28, R=1, S=1 & 51,380,224  \\
\hline
ResNet50-L6  & K=256, C=256, Y=14, X=14, R=3, S=3 & 115,605,504   \\
\hline
BERT-L1 & M=512, N=768, K=768 & 301,989,888   \\
\hline
BERT-L2 & M=512, N=512, K=768 & 201,326,592  \\
\hline
BERT-L3 & M=512, N=768, K=512 & 201,326,592   \\
\hline

GPT-L1 & M=256, N=256, K=2,048 &  134,217,728  \\
\hline
GPT-L2 & M=512, N=512, K=2,048 &  536,870,912 \\
\hline
GPT-L3 & M=256, N=256, K=12,288 & 805,306,368   \\
\hline

\end{tabular}
\label{table:dnn-layers}
\vspace{-2em}
\end{table}
\begin{figure*}[t]
    \centering
    \includegraphics[width=0.95\linewidth]{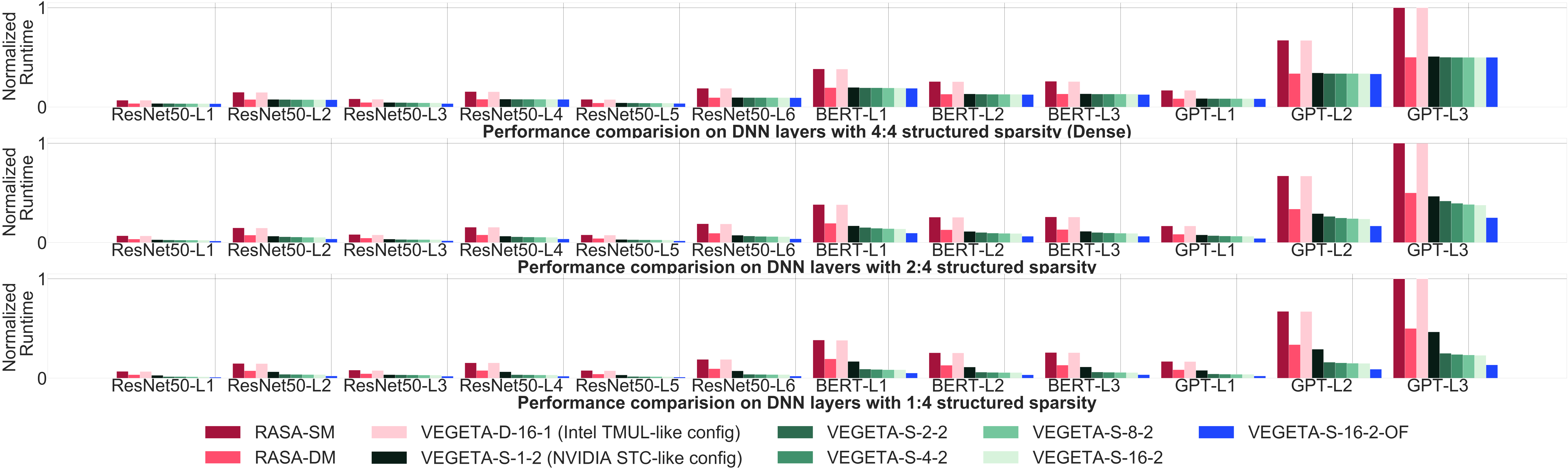}
    \vspace{-1em}
    \caption{
    Normalized runtime with different matrix engines in~\autoref{table:vegeta_designs}.
    We use reddish colors, black, greenish colors, and blue for dense matrix engines~\cite{rasa_dac21, intel2020isa}, a design using STC-like config, config~\cite{mishra2021accelerating},  VEGETA-S designs, and VEGETA-S with OF, respectively.
    }
    \vspace{-1.5em}
    \label{fig:dense_sparse_comparison}
\end{figure*}

\begin{figure}[t]
    \centering
    \includegraphics[width=0.96\linewidth]{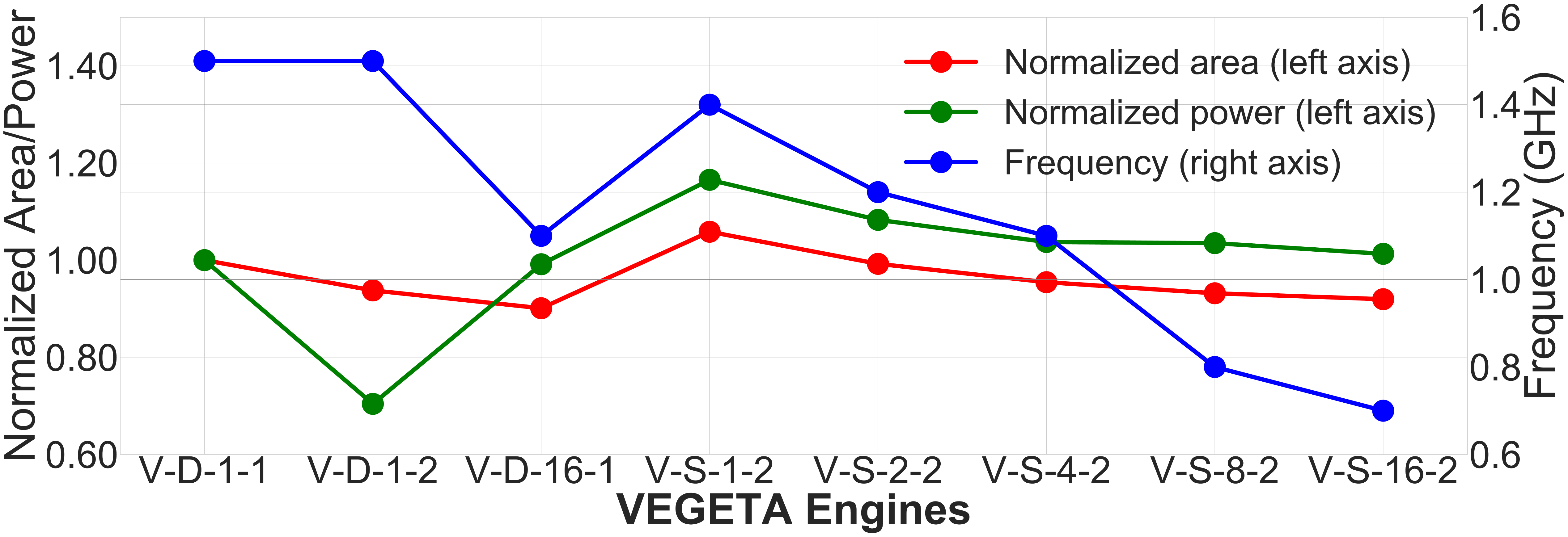}
    \vspace{-1em}
    \caption{
    Area and power normalized to RASA-SM and frequency for different VEGETA engines. V indicates VEGETA.
    }
    \label{fig:area_freq}
    \vspace{-1em}
\end{figure}

\subsection{VEGETA Implementation}
We modified LLVM~\cite{llvm} for the new VEGETA ISA and implemented VEGETA C++ intrinsics.
Next, we wrote GEMM/SPMM kernels that exploit layer-wise $N$:$M$ sparsity using VEGETA intrinsics.
Since there is no commercial CPU that can execute VEGETA instructions,
we developed a Pintool, an instrumentation tool using Pin APIs~\cite{pin_pldi05}, which registers instrumentation callback routines that emulate each of the VEGETA instructions described in~\autoref{table:amx_instructions}.
Then, we generated the traces of the kernels which have every executed instruction with dynamic information using the Pintool, and extended MacSim~\cite{hyesoon12macsim} 
to handle VEGETA instructions and registers along with the different executions of matrix engines.
Finally, we simulated the GEMM/SPMM kernels using the generated traces on MacSim.

We also developed RTL designs to explore different VEGETA engines with different $\alpha$ and $\beta$. 
We model baseline dense matrix engines with RASA-SM, RASA-DM, and the Intel TMUL-like config through VEGETA-D-1-1, VEGETA-D-1-2, and VEGETA-D-16-1, respectively.
We estimate the performance of an engine with the NVIDIA STC-like config using VEGETA-S-1-2 forcing only 2:4 support.
We synthesized each RTL design using Synopsis DC compiler with Nangate 15nm Technology Library. 
We used the post-layout design for area/power/timing numbers for designs shown in~\autoref{table:vegeta_designs}.  

\subsection{Experimental Setup}
Although VEGETA is not limited to a single use case, thanks to its generic SPMM instructions that can be used in various kernels, we use DL inferences as a use case to show the performance of VEGETA.
As DNN compression is done offline (usually once before deployment) for inference~\cite{zhou2021learning, domino2021nips, maohua_micro19, nvidia_ampere}, the cost of DNN compression is amortized to multiple inferences, thus the inference performance does not include this.
For the workload, we choose representative DNN layers from ResNet50~\cite{resnet_cvpr16}, BERT~\cite{devlin-etal-2019-bert}, and GPT-3~\cite{GPT}.
The parameters for the layers are summarized in~\autoref{table:dnn-layers}.
To convert convolutional layers for ResNet50 to GEMM kernels, we use the dimensions derived by applying image to column (im2col) algorithm.
We run the DNN layers with 1:4/2:4/4:4 structured sparsity on different VEGETA engines listed in~\autoref{table:vegeta_designs} using MacSim.
We set the frequency of the core as 2 GHz and fetch/issue/retire width as four with 16 pipeline
stages, 97 ROB entries, and 96 load buffer entries. 
To focus on the performance trade-off of different VEGETA designs, we assume that the data is prefetched to the L2 cache.

\subsection{Performance Analysis}
\autoref{fig:dense_sparse_comparison} shows the runtime with various DNN layers. 
For this experiment, we assume that all the matrix engines are running with 0.5 GHz.
We chose 0.5 GHz since it met the timing for all matrix designs that we used in the evaluation, which are derived based on the corresponding RTL implementations.
We normalized the runtime using the longest runtime (runtime on GPT-L3 with RASA-SM).
We first observe that the RASA-SM suffers from the under-utilization of processing elements due to the mismatch of matrix engine pipeline stages (WL/FF/FS/DR), resulting in the highest runtime.
RASA-DM is a state-of-the-art matrix engine for CPUs and achieves good throughput by matching the latencies of its matrix engine pipeline stages.
Compared to RASA-DM, our sparse engine designs performs comparably for the dense workload showing a performance gain of up to 7\%.
The performance gain is mainly coming from the reduced latency of the drain stage by reducing the width of the SA and output forwarding.

Since VEGETA-D engines are not able to leverage sparsity, they cannot skip ineffectual computations and show the same performance with 2:4 and 1:4 structured sparsity, unlike VEGETA-S engines.
Also, the design with the STC-like config can only accelerate 2:4 sparsity while our VEGETA-S designs can accelerate various fine-grained structured sparsities.
\rev{
For layers with 2:4 structured sparsity, the matrix engine using the STC-like config shows 16\% runtime reduction on average compared with the RASA-DM.
Using VEGETA-S-16-2, additional 18\% runtime reduction was achieved compared to the design with the STC-like config.
Finally, with output forwarding, another 32\% runtime was reduced.}
For layers with 1:4 structured sparsity, the design with the STC-like config does not show better performance compared with 2:4 structured sparsity since it cannot exploit the extra zeros to skip extra ineffectual computations, unlike our VEGETA-S designs.
We observe that VEGETA-S-1-2 shows 51\% runtime reduction on average compared with the RASA-DM since it can skip all ineffectual computations.
Using our VEGETA-S-16-2 engine, additional 8\% runtime reduction was achieved.
Finally, with output forwarding, another 37\% runtime was reduced by resolving output dependencies early.

\subsection{Area and Power Analysis}
\begin{figure}[t]
    \centering
    \includegraphics[width=0.99\linewidth]{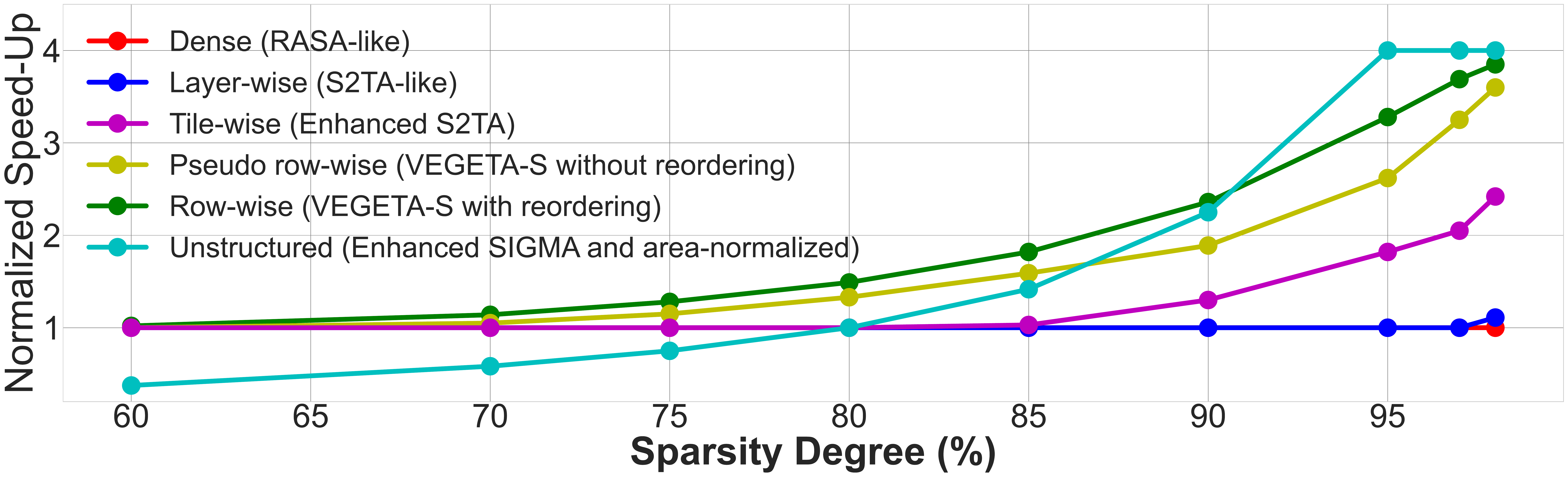}
    \vspace{-2em}
    \caption{
    Average of normalized speed-ups of different sparsity granularity HW support and sparsity degrees using workloads in \autoref{table:dnn-layers}.
    }
    \label{fig:row-wise-eval}
    \vspace{-1em}
\end{figure}
In~\autoref{fig:area_freq}, we show the normalized area and frequency for different VEGETA engines. 
First, we observe that when we increase the number of PUs in a PE ($\alpha$), the area of the VEGETA engines decreases due to the lower number of horizontal pipeline buffers as described in~\autoref{subsec:pu-pe}.
Since we add small modules to support sparsity, the VEGETA-S design with the largest area overhead compared with RASA-SM only causes 6\% area overhead. 
Moreover, by increasing $\alpha$, VEGETA-S-8-2 and VEGETA-S-16-2 show lower area compared to RASA-SM or state-of-the-art dense matrix engine for CPU (RASA-DM).
This is because the overhead gets amortized and compensated as more MACs share the data, reducing the total pipeline registers.
Power shows a similar trend. When we vary $\alpha$ for VEGETA-S-$\alpha$-2 as 1, 2, 4, 8, 16, the power overhead (both static and dynamic) is 17\%, 8\%, 4\%, 3\%, 1\% compared with the baseline.
In the meantime, higher $\alpha$ limits maximum frequency due to the increased wire length for broadcasting across PUs.

\subsection{Analysis for Unstructured Sparsity Support Using VEGETA} \label{sec:row-wise-evalution}

We convert unstructured sparse matrices into row-wise $N$:4 sparse matrices to accelerate SPMM with VEGETA, as discussed in \autoref{subsec:row-wise-support}.
Since there is no work on CPU sparse matrix engines, we use a few SOTA sparse accelerators~\cite{sigma,s2ta_hpca22} as baseline matrix engines for comparison.
As shown in \autoref{table:nm-hw}, S2TA~\cite{s2ta_hpca22} naturally supports layer-wise $N$:$M$ and potentially be enhanced to support tile-wise $N$:$M$ while VEGETA can support pseudo row-wise and row-wise $N$:$M$.
SIGMA~\cite{sigma} can leverage unstructured sparsity, but it comes with area overhead (we normalize performance with the area).
For the conservative evaluation, we assume that they are also enhanced to fully hide fill and drain overhead through perfect pipelining.
Unlike a layer-wise $N$:4 evaluation, it is not straightforward to implement optimized kernels using VEGETA instructions, so we use an analytical roofline model, leaving the development of optimized SW kernels using VEGETA instructions as future work. 
We use the same workloads, but induce random and unstructured sparsity of varying degrees and report average speed-up normalized to a dense engine in \autoref{fig:row-wise-eval}. 
A smaller sparsity granularity increases the possibility of finding $N$:$M$ sparsity that covers non-zeros.  For example, 
it is unlikely that an entire unstructured sparse layer exhibits a certain $N$:$M$ sparsity; thus, layer-wise does not show much performance improvement over dense.
In contrast, row-wise achieves 2.36$\times$ and 3.28$\times$ at 90\% and 95\% sparsity degree.
SIGMA performs better than others with extremely high sparsity degrees ($>$95\%), but it is inefficient for the modest sparsity degree (the target of our work) indicating that extra area overhead is not useful.
\vspace{0em}



\section{Related Work}

\textbf{CPU support to run GEMMs efficiently.} SAVE~\cite{save_micro20} is a sparsity-aware CPU vector engine that skips redundant computations in sparse GEMM operations to accelerate sparse DNN and high-performance computing workloads. Similarly, SparCE~\cite{sparce_taco19} also increases the utilization of vector engines by tracking the general-purpose registers with zeros.
With extremely high sparsity, a program gets memory bounded due to the low arithmetic intensity, making
vector compute throughput enough, and vector engines like SAVE/SparCE can be equal performance to a matrix engine. Otherwise, we expect them to be significantly slower than VEGETA due to the lower compute throughput.
ZCOMP~\cite{zcomp_micro19} introduces a vector ISA extension to reduce cross-layer communication for DNNs.
Our work is orthogonal and complementary to those works since we are targeting a matrix engine that operates on tiles of the matrix instead of individual vectors.
RASA~\cite{rasa_dac21} proposes control and data optimizations for CPU matrix engines to improve utilization 
via efficient pipelining and overlap.
They divide a matrix multiplication with different sub-stages on the systolic array and introduce optimizations with pipelining and overlapping different stages.
It inspired a lot on our work regarding pipelining with sub-stages. However, it does not consider SPMM and their design cannot be used directly as a sparse matrix engine.

\textbf{Handling dynamic sparsity.}
Since dynamic (input) sparsity is hard to predict, it has to be handled at runtime, unlike static sparsity which is usually pre-determined. 
We could use compaction of tile registers to build non-zero tiles, similar to the approach that SAVE~\cite{save_micro20} used for merging vector registers to remove zeros. However, this is not practical for a matrix engine due to the high probability of conflicts across different tiles since the number of operands in a vector register is 32 while that of a tile register is 512 (16$\times$32).
There could be an efficient way 
to exploit dynamic sparsity on matrix engines in CPUs 
without much overhead, but we leave it as future work.




\textbf{Sparse DNN accelerators.} There have been several papers focusing on SPMM/SPGEMM acceleration with standalone accelerators~\cite{scnn, sigma, extensor, srivastava2020matraptor, srivastava2020tensaurus, outerspace, zhang2020sparch}.
Sparse-TPU~\cite{sparsetpu_ics20} and the work from Kung et al.~\cite{columncombining_asplos19} introduce packing algorithms targeting unstructured sparsity for systolic array-like architectures.
Being standalone accelerators with large area-power budgets, these works have explored co-design of compression formats, dataflow, indexing logic, and control structures. 
CPU matrix engines, on the other hand, have a tight coupling with the RF and operate in strict area-power budgets, motivating our solution of enhancing a dense systolic array with a pipeline-friendly dataflow.
Zhu et al.~\cite{maohua_micro19} introduce a sparse tensor core by extending NVIDIA Tensor Core using vector-wise pruning and encoding to enforce structured sparsity. 
NVIDIA has also recently introduced Sparse Tensor Core as a sparse matrix engine in their Ampere GPUs~\cite{mishra2021accelerating} to accelerate sparse DNNs~\cite{mishra2021accelerating} with 2:4 structured sparsity. 
STA~\cite{sta_cal20} proposes a systolic tensor array for inference on mobile devices with density-bound block sparsity, which is similar to the $N$:$M$ structured sparsity, and S2TA~\cite{s2ta_hpca22} extended STA for flexible $N$:$M$ sparsity.
None of the previous works supports row-wise $N$:$M$ sparsity, which this work shows as a promising technique to cover unstructured sparsity. Also, prior works do not consider dividing the execution of tile computations into fine-grained stages for pipelining without resource conflicts, which is critical when integrated into CPUs.

\section{Conclusion}
This work adds flexible $N$:$M$ structured sparsity support in CPU matrix engines via VEGETA ISA and engines supporting various sparsity granularity.   
We propose several design choices for the VEGETA engine, studying kernel performance, clock frequency, and area trade-offs.
Also, we show how efficiently the row-wise $N$:$M$ sparsity feature of VEGETA can accelerate unstructured sparsity by transforming unstructured sparsity to row-wise $N$:$M$ sparsity.
We believe this work opens up opportunities for future work in enhancing the design further for dynamic sparsity and studying the interaction of the sparse matrix engine with the rest of the CPU memory hierarchy.



\section{Acknowledgement}
This work was funded by an award from Intel’s Corporate Research Council.
We thank Raveesh Garg, Po-An Tsai, and Vivek Sarkar for profound discussions on this work. 

\newpage

\bibliographystyle{hpca23-templete/IEEEtranS}
\bibliography{hpca23-templete/reference}

\begin{thebibliography}{10}
\providecommand{\url}[1]{#1}
\csname url@samestyle\endcsname
\providecommand{\newblock}{\relax}
\providecommand{\bibinfo}[2]{#2}
\providecommand{\BIBentrySTDinterwordspacing}{\spaceskip=0pt\relax}
\providecommand{\BIBentryALTinterwordstretchfactor}{4}
\providecommand{\BIBentryALTinterwordspacing}{\spaceskip=\fontdimen2\font plus
\BIBentryALTinterwordstretchfactor\fontdimen3\font minus
  \fontdimen4\font\relax}
\providecommand{\BIBforeignlanguage}[2]{{%
\expandafter\ifx\csname l@#1\endcsname\relax
\typeout{** WARNING: IEEEtranS.bst: No hyphenation pattern has been}%
\typeout{** loaded for the language `#1'. Using the pattern for}%
\typeout{** the default language instead.}%
\else
\language=\csname l@#1\endcsname
\fi
#2}}
\providecommand{\BIBdecl}{\relax}
\BIBdecl

\bibitem{gemm_petewarden}
``https://petewarden.com/2015/04/20/why-gemm-is-at-the-heart-of-deep-learning/,''
  2015.

\bibitem{nvidia_volta}
``Nvidia tesla v100 gpu architecture.'' 2017,
  \url{https://images.nvidia.com/content/volta-architecture/pdf/volta-architecture-whitepaper.pdf}.

\bibitem{nvidia-bf16}
``Tensorfloat-32 in the a100 gpu accelerates ai training, hpc up to 20x,''
  2019,
  \url{https://blogs.nvidia.com/blog/2020/05/14/tensorfloat-32-precision-format/}.

\bibitem{nvidia_ampere}
``Nvidia ampere ga102 gpu architecture.'' 2021,
  \url{https://www.nvidia.com/content/PDF/nvidia-ampere-ga-102-gpu-architecture-whitepaper-v2.pdf}.

\bibitem{onednn}
``oneapi deep neural network library (onednn),'' 2021,
  \url{https://github.com/oneapi-src/oneDNN}.

\bibitem{sambanova}
\BIBentryALTinterwordspacing
``Sambanova whitepaper,'' 2021. [Online]. Available:
  \url{https://sambanova.ai/wp-content/uploads/2021/06/SambaNova_RDA_Whitepaper_English.pdf}
\BIBentrySTDinterwordspacing

\bibitem{tensorflow2015-whitepaper}
\BIBentryALTinterwordspacing
M.~Abadi, A.~Agarwal, P.~Barham, E.~Brevdo, Z.~Chen, C.~Citro, G.~S. Corrado,
  A.~Davis, J.~Dean, M.~Devin, S.~Ghemawat, I.~Goodfellow, A.~Harp, G.~Irving,
  M.~Isard, Y.~Jia, R.~Jozefowicz, L.~Kaiser, M.~Kudlur, J.~Levenberg,
  D.~Man\'{e}, R.~Monga, S.~Moore, D.~Murray, C.~Olah, M.~Schuster, J.~Shlens,
  B.~Steiner, I.~Sutskever, K.~Talwar, P.~Tucker, V.~Vanhoucke, V.~Vasudevan,
  F.~Vi\'{e}gas, O.~Vinyals, P.~Warden, M.~Wattenberg, M.~Wicke, Y.~Yu, and
  X.~Zheng, ``{TensorFlow}: Large-scale machine learning on heterogeneous
  systems,'' 2015, software available from tensorflow.org. [Online]. Available:
  \url{https://www.tensorflow.org/}
\BIBentrySTDinterwordspacing

\bibitem{zcomp_micro19}
\BIBentryALTinterwordspacing
B.~Akin, Z.~A. Chishti, and A.~R. Alameldeen, ``Zcomp: Reducing dnn cross-layer
  memory footprint using vector extensions,'' in \emph{Proceedings of the 52nd
  Annual IEEE/ACM International Symposium on Microarchitecture}, ser. MICRO
  '52.\hskip 1em plus 0.5em minus 0.4em\relax New York, NY, USA: Association
  for Computing Machinery, 2019, p. 126–138. [Online]. Available:
  \url{https://doi.org/10.1145/3352460.3358305}
\BIBentrySTDinterwordspacing

\bibitem{arm20ethos}
ARM, ``Powering the edge: Driving optimal performance with the ethos-n77 npu,''
  \emph{Whitepaper}, 2019.

\bibitem{GPT}
\BIBentryALTinterwordspacing
T.~B. Brown, B.~Mann, N.~Ryder, M.~Subbiah, J.~Kaplan, P.~Dhariwal,
  A.~Neelakantan, P.~Shyam, G.~Sastry, A.~Askell, S.~Agarwal, A.~Herbert-Voss,
  G.~Krueger, T.~Henighan, R.~Child, A.~Ramesh, D.~M. Ziegler, J.~Wu,
  C.~Winter, C.~Hesse, M.~Chen, E.~Sigler, M.~Litwin, S.~Gray, B.~Chess,
  J.~Clark, C.~Berner, S.~McCandlish, A.~Radford, I.~Sutskever, and D.~Amodei,
  ``Language models are few-shot learners,'' 2020. [Online]. Available:
  \url{https://arxiv.org/abs/2005.14165}
\BIBentrySTDinterwordspacing

\bibitem{tvm}
T.~Chen, T.~Moreau, Z.~Jiang, L.~Zheng, E.~Yan, M.~Cowan, H.~Shen, L.~Wang,
  Y.~Hu, L.~Ceze, C.~Guestrin, and A.~Krishnamurthy, ``Tvm: An automated
  end-to-end optimizing compiler for deep learning,'' in \emph{Proceedings of
  the 13th USENIX Conference on Operating Systems Design and Implementation},
  ser. OSDI'18.\hskip 1em plus 0.5em minus 0.4em\relax USA: USENIX Association,
  2018, p. 579–594.

\bibitem{eyeriss}
Y.-H. Chen, T.~Krishna, J.~S. Emer, and V.~Sze, ``Eyeriss: An energy-efficient
  reconfigurable accelerator for deep convolutional neural networks,''
  \emph{IEEE Journal of Solid-State Circuits}, vol.~52, no.~1, pp. 127--138,
  2017.

\bibitem{eyeriss_v2}
Y.-H. Chen, T.-J. Yang, J.~Emer, and V.~Sze, ``Eyeriss v2: A flexible
  accelerator for emerging deep neural networks on mobile devices,'' \emph{IEEE
  Journal on Emerging and Selected Topics in Circuits and Systems}, vol.~9,
  no.~2, pp. 292--308, 2019.

\bibitem{cudnn}
S.~Chetlur, C.~Woolley, P.~Vandermersch, J.~Cohen, J.~Tran, B.~Catanzaro, and
  E.~Shelhamer, ``cudnn: Efficient primitives for deep learning,'' \emph{arXiv
  preprint arXiv:1410.0759}, 2014.

\bibitem{devlin-etal-2019-bert}
\BIBentryALTinterwordspacing
J.~Devlin, M.-W. Chang, K.~Lee, and K.~Toutanova, ``{BERT}: Pre-training of
  deep bidirectional transformers for language understanding,'' in
  \emph{Proceedings of the 2019 Conference of the North {A}merican Chapter of
  the Association for Computational Linguistics: Human Language Technologies,
  Volume 1 (Long and Short Papers)}.\hskip 1em plus 0.5em minus 0.4em\relax
  Minneapolis, Minnesota: Association for Computational Linguistics, Jun. 2019,
  pp. 4171--4186. [Online]. Available: \url{https://aclanthology.org/N19-1423}
\BIBentrySTDinterwordspacing

\bibitem{geng2019awb}
T.~Geng, A.~Li, T.~Wang, C.~Wu, Y.~Li, R.~Shi, A.~Tumeo, S.~Che, S.~Reinhardt,
  and M.~Herbordt, ``Awb-gcn: A graph convolutional network accelerator with
  runtime workload rebalancing,'' \emph{MICRO}, 2020.

\bibitem{libxsmm_ipdps20}
E.~Georganas, K.~Banerjee, D.~Kalamkar, S.~Avancha, A.~Venkat, M.~Anderson,
  G.~Henry, H.~Pabst, and A.~Heinecke, ``Harnessing deep learning via a single
  building block,'' in \emph{2020 IEEE International Parallel and Distributed
  Processing Symposium (IPDPS)}, 2020, pp. 222--233.

\bibitem{save_micro20}
Z.~Gong, H.~Ji, C.~W. Fletcher, C.~J. Hughes, S.~Baghsorkhi, and J.~Torrellas,
  ``Save: Sparsity-aware vector engine for accelerating dnn training and
  inference on cpus,'' in \emph{MICRO}, 2020.

\bibitem{deepcompression_iclr16}
S.~Han, H.~Mao, and W.~J. Dally, ``Deep compression: Compressing deep neural
  network with pruning, trained quantization and huffman coding,'' in \emph{4th
  International Conference on Learning Representations, {ICLR} 2016, San Juan,
  Puerto Rico, May 2-4, 2016, Conference Track Proceedings}, Y.~Bengio and
  Y.~LeCun, Eds., 2016.

\bibitem{pruning_nips15}
S.~Han, J.~Pool, J.~Tran, and W.~J. Dally, ``Learning both weights and
  connections for efficient neural networks,'' in \emph{Proceedings of the 28th
  International Conference on Neural Information Processing Systems - Volume
  1}, ser. NIPS'15.\hskip 1em plus 0.5em minus 0.4em\relax Cambridge, MA, USA:
  MIT Press, 2015, p. 1135–1143.

\bibitem{resnet_cvpr16}
K.~He, X.~Zhang, S.~Ren, and J.~Sun, ``Deep residual learning for image
  recognition,'' in \emph{2016 IEEE Conference on Computer Vision and Pattern
  Recognition (CVPR)}, 2016, pp. 770--778.

\bibitem{sparsetpu_ics20}
\BIBentryALTinterwordspacing
X.~He, S.~Pal, A.~Amarnath, S.~Feng, D.-H. Park, A.~Rovinski, H.~Ye, Y.~Chen,
  R.~Dreslinski, and T.~Mudge, ``Sparse-tpu: Adapting systolic arrays for
  sparse matrices,'' in \emph{Proceedings of the 34th ACM International
  Conference on Supercomputing}, ser. ICS '20.\hskip 1em plus 0.5em minus
  0.4em\relax New York, NY, USA: Association for Computing Machinery, 2020.
  [Online]. Available: \url{https://doi.org/10.1145/3392717.3392751}
\BIBentrySTDinterwordspacing

\bibitem{extensor}
\BIBentryALTinterwordspacing
K.~Hegde, H.~Asghari-Moghaddam, M.~Pellauer, N.~Crago, A.~Jaleel, E.~Solomonik,
  J.~Emer, and C.~W. Fletcher, ``Extensor: An accelerator for sparse tensor
  algebra,'' in \emph{Proceedings of the 52nd Annual IEEE/ACM International
  Symposium on Microarchitecture}, ser. MICRO '52.\hskip 1em plus 0.5em minus
  0.4em\relax New York, NY, USA: Association for Computing Machinery, 2019, p.
  319–333. [Online]. Available: \url{https://doi.org/10.1145/3352460.3358275}
\BIBentrySTDinterwordspacing

\bibitem{ibm20isa}
IBM, ``Power isa version 3.1,'' \emph{Whitepaper}, 2020.

\bibitem{intelBWcal_orig}
Intel, ``Theoretical maximum memory bandwidth for intel® core™ x-series
  processors,''
  \url{https://www.intel.com/content/www/us/en/support/articles/000056722/processors/intel-core-processors.html}.

\bibitem{intel2020isa}
Intel, ``Intel 64 and ia-32 architectures software developer’s manual,''
  2020.

\bibitem{intel2021}
Intel, ``Presentation deck: Intel architecture day 2021,'' 2021,
  \url{https://download.intel.com/newsroom/2021/client-computing/intel-architecture-day-2021-presentation.pdf}.

\bibitem{samsung}
J.-W. Jang, S.~Lee, D.~Kim, H.~Park, A.~S. Ardestani, Y.~Choi, C.~Kim, Y.~Kim,
  H.~Yu, H.~Abdel-Aziz, J.-S. Park, H.~Lee, D.~Lee, M.~W. Kim, H.~Jung, H.~Nam,
  D.~Lim, S.~Lee, J.-H. Song, S.~Kwon, J.~Hassoun, S.~Lim, and C.~Choi,
  ``Sparsity-aware and re-configurable npu architecture for samsung flagship
  mobile soc,'' in \emph{2021 ACM/IEEE 48th Annual International Symposium on
  Computer Architecture (ISCA)}, 2021, pp. 15--28.

\bibitem{rasa_dac21}
G.~Jeong, E.~Qin, A.~Samajdar, C.~J. Hughes, S.~Subramoney, H.~Kim, and
  T.~Krishna, ``Rasa: Efficient register-aware systolic array matrix engine for
  cpu,'' in \emph{2021 58th ACM/IEEE Design Automation Conference (DAC)}, 2021,
  pp. 253--258.

\bibitem{tpu-isca}
N.~P. Jouppi, , C.~Young, N.~Patil, D.~Patterson, G.~Agrawal, R.~Bajwa,
  S.~Bates, S.~Bhatia, N.~Boden, A.~Borchers, R.~Boyle, P.~l. Cantin, C.~Chao,
  C.~Clark, J.~Coriell, M.~Daley, M.~Dau, J.~Dean, B.~Gelb, T.~V. Ghaemmaghami,
  R.~Gottipati, W.~Gulland, R.~Hagmann, C.~R. Ho, D.~Hogberg, J.~Hu, R.~Hundt,
  D.~Hurt, J.~Ibarz, A.~Jaffey, A.~Jaworski, A.~Kaplan, H.~Khaitan,
  D.~Killebrew, A.~Koch, N.~Kumar, S.~Lacy, J.~Laudon, J.~Law, D.~Le, C.~Leary,
  Z.~Liu, K.~Lucke, A.~Lundin, G.~MacKean, A.~Maggiore, M.~Mahony, K.~Miller,
  R.~Nagarajan, R.~Narayanaswami, R.~Ni, K.~Nix, T.~Norrie, M.~Omernick,
  N.~Penukonda, A.~Phelps, J.~Ross, M.~Ross, A.~Salek, E.~Samadiani, C.~Severn,
  G.~Sizikov, M.~Snelham, J.~Souter, D.~Steinberg, A.~Swing, M.~Tan,
  G.~Thorson, B.~Tian, H.~Toma, E.~Tuttle, V.~Vasudevan, R.~Walter, W.~Wang,
  E.~Wilcox, and D.~H. Yoon, ``In-datacenter performance analysis of a tensor
  processing unit,'' in \emph{Proceedings of the 44th Annual International
  Symposium on Computer Architecture (ISCA)}, 2017.

\bibitem{hyesoon12macsim}
H.~Kim \emph{et~al.}, ``Macsim: A cpu-gpu heterogeneous simulation framework
  user guide,'' 2012.

\bibitem{intel-mlperf}
P.~D. Koichi~Yamada, Wei~Li, ``Intel’s mlperf results show robust cpu-based
  training performance for a range of workloads,'' 2020,
  \url{https://www.intel.com/content/www/us/en/artificial-intelligence/posts/intels-mlper-results.html}.

\bibitem{systolic}
H.-T. Kung, ``Why systolic architectures?'' \emph{Computer}, no.~1, pp. 37--46,
  1982.

\bibitem{columncombining_asplos19}
\BIBentryALTinterwordspacing
H.~Kung, B.~McDanel, and S.~Q. Zhang, ``Packing sparse convolutional neural
  networks for efficient systolic array implementations: Column combining under
  joint optimization,'' in \emph{Proceedings of the Twenty-Fourth International
  Conference on Architectural Support for Programming Languages and Operating
  Systems}, ser. ASPLOS '19.\hskip 1em plus 0.5em minus 0.4em\relax New York,
  NY, USA: Association for Computing Machinery, 2019, p. 821–834. [Online].
  Available: \url{https://doi.org/10.1145/3297858.3304028}
\BIBentrySTDinterwordspacing

\bibitem{llvm}
C.~Lattner and V.~Adve, ``Llvm: a compilation framework for lifelong program
  analysis \& transformation,'' in \emph{International Symposium on Code
  Generation and Optimization, 2004. CGO 2004.}, 2004, pp. 75--86.

\bibitem{sta_cal20}
Z.-G. Liu, P.~N. Whatmough, and M.~Mattina, ``Systolic tensor array: An
  efficient structured-sparse gemm accelerator for mobile cnn inference,''
  \emph{IEEE Computer Architecture Letters}, vol.~19, no.~1, pp. 34--37, 2020.

\bibitem{s2ta_hpca22}
Z.-G. Liu, P.~N. Whatmough, Y.~Zhu, and M.~Mattina, ``S2ta: Exploiting
  structured sparsity for energy-efficient mobile cnn acceleration,'' in
  \emph{2022 IEEE International Symposium on High-Performance Computer
  Architecture (HPCA)}, 2022, pp. 573--586.

\bibitem{pin_pldi05}
\BIBentryALTinterwordspacing
C.-K. Luk, R.~Cohn, R.~Muth, H.~Patil, A.~Klauser, G.~Lowney, S.~Wallace, V.~J.
  Reddi, and K.~Hazelwood, ``Pin: Building customized program analysis tools
  with dynamic instrumentation,'' \emph{SIGPLAN Not.}, vol.~40, no.~6, p.
  190–200, jun 2005. [Online]. Available:
  \url{https://doi.org/10.1145/1064978.1065034}
\BIBentrySTDinterwordspacing

\bibitem{mishra2021accelerating}
A.~Mishra, J.~A. Latorre, J.~Pool, D.~Stosic, D.~Stosic, G.~Venkatesh, C.~Yu,
  and P.~Micikevicius, ``Accelerating sparse deep neural networks,''
  \emph{arXiv preprint arXiv:2104.08378}, 2021.

\bibitem{dlrm19}
\BIBentryALTinterwordspacing
M.~Naumov, D.~Mudigere, H.~M. Shi, J.~Huang, N.~Sundaraman, J.~Park, X.~Wang,
  U.~Gupta, C.~Wu, A.~G. Azzolini, D.~Dzhulgakov, A.~Mallevich,
  I.~Cherniavskii, Y.~Lu, R.~Krishnamoorthi, A.~Yu, V.~Kondratenko, S.~Pereira,
  X.~Chen, W.~Chen, V.~Rao, B.~Jia, L.~Xiong, and M.~Smelyanskiy, ``Deep
  learning recommendation model for personalization and recommendation
  systems,'' \emph{CoRR}, vol. abs/1906.00091, 2019. [Online]. Available:
  \url{https://arxiv.org/abs/1906.00091}
\BIBentrySTDinterwordspacing

\bibitem{outerspace}
S.~Pal, J.~Beaumont, D.-H. Park, A.~Amarnath, S.~Feng, C.~Chakrabarti, H.-S.
  Kim, D.~Blaauw, T.~Mudge, and R.~Dreslinski, ``Outerspace: An outer product
  based sparse matrix multiplication accelerator,'' in \emph{ISCA}, 2018.

\bibitem{scnn}
A.~Parashar, M.~Rhu, A.~Mukkara, A.~Puglielli, R.~Venkatesan, B.~Khailany,
  J.~Emer, S.~W. Keckler, and W.~J. Dally, ``Scnn: An accelerator for
  compressed-sparse convolutional neural networks,'' in \emph{2017 ACM/IEEE
  44th Annual International Symposium on Computer Architecture (ISCA)}, 2017,
  pp. 27--40.

\bibitem{sigma}
E.~Qin, A.~Samajdar, H.~Kwon, V.~Nadella, S.~Srinivasan, D.~Das, B.~Kaul, and
  T.~Krishna, ``Sigma: A sparse and irregular gemm accelerator with flexible
  interconnects for dnn training,'' in \emph{2020 IEEE International Symposium
  on High Performance Computer Architecture (HPCA)}.\hskip 1em plus 0.5em minus
  0.4em\relax IEEE, 2020, pp. 58--70.

\bibitem{cerebras}
K.~Rocki, D.~Van~Essendelft, I.~Sharapov, R.~Schreiber, M.~Morrison,
  V.~Kibardin, A.~Portnoy, J.~F. Dietiker, M.~Syamlal, and M.~James, ``Fast
  stencil-code computation on a wafer-scale processor,'' in \emph{SC20:
  International Conference for High Performance Computing, Networking, Storage
  and Analysis}.\hskip 1em plus 0.5em minus 0.4em\relax IEEE, 2020, pp. 1--14.

\bibitem{intel2018vnni}
A.~Rodrigues \emph{et~al.}, ``Lower numerical precision deep learning inference
  and training,'' \emph{Whitepaper}, 2018.

\bibitem{sparce_taco19}
\BIBentryALTinterwordspacing
S.~Sen, S.~Jain, S.~Venkataramani, and A.~Raghunathan, ``Sparce: Sparsity aware
  general-purpose core extensions to accelerate deep neural networks,''
  \emph{IEEE Trans. Comput.}, vol.~68, no.~6, p. 912–925, Jun. 2019.
  [Online]. Available: \url{https://doi.org/10.1109/TC.2018.2879434}
\BIBentrySTDinterwordspacing

\bibitem{tpu-bf16}
P.~K. Shibo~Wang, ``Bfloat16: The secret to high performance on cloud tpus,''
  2019,
  \url{https://cloud.google.com/blog/products/ai-machine-learning/bfloat16-the-secret-to-high-performance-on-cloud-tpus}.

\bibitem{accelerometer}
\BIBentryALTinterwordspacing
A.~Sriraman and A.~Dhanotia, ``Accelerometer: Understanding acceleration
  opportunities for data center overheads at hyperscale,'' in \emph{Proceedings
  of the Twenty-Fifth International Conference on Architectural Support for
  Programming Languages and Operating Systems}, ser. ASPLOS '20.\hskip 1em plus
  0.5em minus 0.4em\relax New York, NY, USA: Association for Computing
  Machinery, 2020, p. 733–750. [Online]. Available:
  \url{https://doi.org/10.1145/3373376.3378450}
\BIBentrySTDinterwordspacing

\bibitem{srivastava2020matraptor}
N.~Srivastava, H.~Jin, J.~Liu, D.~Albonesi, and Z.~Zhang, ``Matraptor: A
  sparse-sparse matrix multiplication accelerator based on row-wise product,''
  in \emph{2020 53rd Annual IEEE/ACM International Symposium on
  Microarchitecture (MICRO)}, 2020, pp. 766--780.

\bibitem{srivastava2020tensaurus}
N.~Srivastava, H.~Jin, S.~Smith, H.~Rong, D.~Albonesi, and Z.~Zhang,
  ``Tensaurus: A versatile accelerator for mixed sparse-dense tensor
  computations,'' in \emph{2020 IEEE International Symposium on High
  Performance Computer Architecture (HPCA)}, 2020, pp. 689--702.

\bibitem{ibm21micro}
W.~J. Starke, B.~W. Thompto, J.~A. Stuecheli, and J.~E. Moreira, ``Ibm's
  power10 processor,'' \emph{IEEE Micro}, vol.~41, no.~2, pp. 7--14, 2021.

\bibitem{domino2021nips}
W.~Sun, A.~Zhou, S.~Stuijk, R.~Wijnhoven, A.~O. Nelson, hongsheng Li, and
  H.~Corporaal, ``Dominosearch: Find layer-wise fine-grained n:m sparse schemes
  from dense neural networks,'' in \emph{Advances in Neural Information
  Processing Systems}, vol.~34, 2021.

\bibitem{mlfb_hpca19}
C.-J. Wu, D.~Brooks, K.~Chen, D.~Chen, S.~Choudhury, M.~Dukhan, K.~Hazelwood,
  E.~Isaac, Y.~Jia, B.~Jia, T.~Leyvand, H.~Lu, Y.~Lu, L.~Qiao, B.~Reagen,
  J.~Spisak, F.~Sun, A.~Tulloch, P.~Vajda, X.~Wang, Y.~Wang, B.~Wasti, Y.~Wu,
  R.~Xian, S.~Yoo, and P.~Zhang, ``Machine learning at facebook: Understanding
  inference at the edge,'' in \emph{2019 IEEE International Symposium on High
  Performance Computer Architecture (HPCA)}, 2019, pp. 331--344.

\bibitem{zhang2020sparch}
Z.~Zhang, H.~Wang, S.~Han, and W.~J. Dally, ``Sparch: Efficient architecture
  for sparse matrix multiplication,'' in \emph{2020 IEEE International
  Symposium on High Performance Computer Architecture (HPCA)}, 2020, pp.
  261--274.

\bibitem{zhou2021learning}
\BIBentryALTinterwordspacing
A.~Zhou, Y.~Ma, J.~Zhu, J.~Liu, Z.~Zhang, K.~Yuan, W.~Sun, and H.~Li,
  ``Learning n:m fine-grained structured sparse neural networks from scratch,''
  in \emph{International Conference on Learning Representations}, 2021.
  [Online]. Available: \url{https://openreview.net/forum?id=K9bw7vqp_s}
\BIBentrySTDinterwordspacing

\bibitem{maohua_micro19}
M.~Zhu, T.~Zhang, Z.~Gu, and Y.~Xie, ``Sparse tensor core: Algorithm and
  hardware co-design for vector-wise sparse neural networks on modern gpus,''
  in \emph{MICRO}, 2019.

\end{thebibliography}

\end{document}